\documentclass[12pt,preprint]{aastex}

\def\lea{\mathrel{<\kern-1.0em\lower0.9ex\hbox{$\sim$}}}
\def\gea{\mathrel{>\kern-1.0em\lower0.9ex\hbox{$\sim$}}}

\slugcomment{Submitted to The Astrophysical Journal}

\shorttitle{Photometric Properties of the M33 Star Cluster System}
\shortauthors{San Roman et al.}

\begin{document}

\title{Photometric Properties of the M33 Star Cluster System
\footnote{Based on observations obtained with MegaPrime/MegaCam, a joint project of CFHT and CEA/DAPNIA, at the Canada-France-Hawaii Telescope (CFHT) which is operated by the National Research Council (NRC) of Canada, the Institut National des Science de l'Univers of the Centre National de la Recherche Scientifique (CNRS) of France, and the University of Hawaii.}}

\author{Izaskun San Roman and Ata Sarajedini}
\affil{Department of Astronomy, University of Florida, \\211 Bryant Space
Science Center, Gainesville, FL 32611-2055}
\email{izaskun@astro.ufl.edu, ata@astro.ufl.edu}

\author{Antonio Aparicio}
\affil{Instituto de Astrof\'\i sica de Canarias, \\V\'\i a L\'actea s/n.
E38205 - La Laguna, Tenerife, Canary Islands, Spain.}
\email{antapaj@iac.es}

\begin{abstract}
We present a catalog of 2,990 extended sources in a $1^{\circ}$x$1^{\circ}$ area centered on M33 using the MegaCam camera on the 3.6m Canada-France-Hawaii telescope (CFHT). The catalog includes 599 new candidate stellar clusters, 204 previously confirmed clusters, 1969 likely background galaxies  and  218 unknown extended objects. We present \textit{ugriz} integrated magnitudes of  the candidates  and confirmed star clusters as well as full width at half maximum, ellipticity and stellarity. Based on the properties of the confirmed star clusters, we select a sub-sample of highly probable clusters composed of 246 objects. The integrated photometry of the complete cluster catalog reveals a wide range of colors from --0.4 $<$ (g--r) $<$ 1.5 and  --1.0 $<$ (r--i) $<$ 1.0 with no obvious cluster subpopulations. Comparisons with models of simple stellar populations suggest a large range of ages some as old as $\sim$10 Gyrs. In addition, we find a sequence in the color-color diagrams that deviates from the expected direction of evolution. This feature could be associated with very young clusters ($<10^{7}$yrs) possessing significant nebular emission. Analysis of the radial density distribution suggests that the cluster system of M33 has suffered from significant depletion possibly due to interactions with M31. We also detect a gap in the cluster distribution in the color-color diagram  at (g--r) $\simeq$ 0.3 and (u--g) $\simeq$ 0.8. This gap could be interpreted as an evolutionary effect. This complete catalog provides promising targets for deep photometry and high resolution spectroscopy to study the structure and star formation history of M33.
\end{abstract}


\section{Introduction}
Star clusters provide important information for understanding the formation and evolution of galaxies. Such systems are useful for highlighting substructures of their host galaxies and for revealing their merging history. In particular, the galaxy formation process can be traced through the ages, metallicities, and kinematics of star clusters. Due to their proximity, galaxies in the Local Group provide us with ideal targets for detailed studies of star cluster properties. While the star cluster systems of the Milky Way and M31 have received close attention, the third spiral galaxy in the Local Group, M33, has been less studied. At a distance of 870 kpc \citep[distance modulus = 24.69;][]{Galletietal2004}, M33 is the only nearby late-type spiral galaxy (Scd). With a large angular size and inclination of i=56$^{\circ}$ \citep{ReganandVogel1994}, M33 is a suitable galaxy for studies of its stellar constituents.\\

There have been a number of M33 cluster catalogs published since the pioneering work of \cite{Hiltner1960}. An extensive and complete catalog can be found in the work of \citet[hereafter SM]{SarajediniandMancone2007}, which merged all of the modern catalogs compiled before 2007. The catalogs that have appeared after the publication of SM have been incorporated into the web-based version of the SM catalog\footnote[2]{http://www.astro.ufl.edu/$\sim$ata/cgi-bin/m33\_cluster\_catalog/index.cgi}. This updated version of the catalog contains 595 candidates of which 349 are confirmed clusters based on \textit{Hubble Space Telescope (HST)} and high-resolution ground-based imaging. The most recent work in this field corresponds to \citet[hereafter ZKH]{Zloczewskietal2008} using the MegaCam camera on CFHT. This study presents a catalog of 4,780 extended sources in a 1 deg$^{2}$ region around M33 which includes 3,554 new candidate stellar clusters.\\

As pointed out by SM, the sample of clusters in M33 suffers from significant incompleteness. While \textit{HST} and its several instruments have been successfully used in the search for star clusters, the small field of view permits surveys only over a limited region of the galaxy. With the most recent contribution of ZKH, this area has been increased to 1 deg$^{2}$ centered on M33.  However, as we discussed in \citet{Sanromanetal2009}, the ZKH catalog has largely overestimated the number of clusters, due to a possible systematic misidentification, where only around 40$\%$ of the 3554 proposed candidates are likely to be actual stellar clusters. For these reasons we have undertaken the present study. This paper is organized as follows: Section 2 describes the observations and data reduction while section 3 discusses the adopted search method and the integrated photometry of the clusters. The analysis of the photometric properties and comparison with other galaxies are in Section 4. Finally, Section 5 presents a summary.
    
\section{Observations and Data Reduction}
                                   
The observations for the present study were obtained using the Queue Service Observing mode at the 3.6 m  Canada France Hawaii Telescope (CFHT). The data are available on-line through \textit{The Canadian Astronomy Data Centre} archive and were obtained as part of ``The M33 CFHT Variability Survey'' \citep{Hartman2006}. The images were taken using the MegaCam/MegaPrime wide-field mosaic imager which contains 36 individual CCDs that combine to offer nearly a full $1^{\circ}$x $1^{\circ}$ field of view with a high angular resolution of 0.187''pixel$^{-1}$. MegaCam operates with a set of g'r'i'z' filters very similar to those of the Sloan Digital Sky Survey (SDSS) but a slightly different near-UV filter called u$^{*}$. This filter was designed to maximize the capabilities of the instrument at short wavelengths and its effective wavelength is $\sim$ 200 $\AA$ redder than the standard u' filter. All of the archival images were pre-processed by the CFHT's Elixir project. This pipeline includes the standard steps of overscan and bias subtraction, flat-fielding, fringe correction, masking of bad pixels and merging of amplifiers. The Elixir project also provides a preliminary photometric calibration for each image.\\

In order to facilitate the search for cluster candidates, only the best available images were analyzed consisting of 15 u$^{*}$, 15 g', 14 r', 28 i' and 3 z'. Median seeing values of all analyzed images are $<$ 0.8'' in g', r' and i' filters and $\sim$ 0.6'' in u$^{*}$ and z'. Prior to the data analysis,  each field of 36 individual CCDs were combined into a single master image. This process was done using the software module Swarp 2.16.14 of the Terapix pipeline which is mainly dedicated to the processing of MegaCam data. This specific module involves resampling of the individual images as well as co-adding the different exposures in an optimum way so that the point-spread function (PSF) is not distorted \citep[for details]{Bertin2002}. Each final combined master image was divided into two sub-fields, including an overlapping area, to deal with the spatial variability of the PSF.\\

The M33 images are extremely crowded making the construction of point-spread functions quite challenging. In order to perform accurate standard profile-fitting photometry, we used DAOPHOT/ALLSTAR routines \citep{Stetson1994} in an iterative way. First, we found all of the stars on each image and produced small-aperture photometry for them. We then used the DAOPHOT/PICK routine to select a set of 1000 reasonable candidates to be used as PSF stars. After deleting those with bad pixels nearby, we subtracted the stars with surrounding neighbors to help isolate the PSF stars. The resulting list of more than 500 stars, in all cases, was used to create a PSF for each of the images. The shape of the PSF was made to vary quadratically with position on the frame. To improve the PSF, we created an image where all the neighbors and stars that do not fit the first PSF were subtracted, obtaining an improved second-generation PSF over this subtracted image. Appropriate aperture corrections were calculated from isolated unsaturated bright stars with photometric errors smaller than 0.01 mag. Since the correction varies with radius from the center of the images, a polynomial fit was applied to the aperture corrections in order to obtain the final instrumental magnitudes. All frames were matched using DAOMATCH/DAOMASTER routines to obtain common stars in all filters. Following \cite{Hartman2006}, the photometric calibration provided by the Elixir pipeline was applied using the zero-point values. In addition and to deal with the differences between u$^{*}$ and u', we applied the equations from \cite{Clem2008} to transform the photometry from $u^{*}g'r'i'z'$ to $u'g'r'i'z'$.\\  

The integrated magnitudes and colors for each candidate cluster have been calculated using the aperture photometry routines in DAOPHOT \citep{Stetson1987}. To be consistent with previous authors \citep{Sanromanetal2009,SarajediniBarker2007,CBF1999,CBF2001}, we have adopted an aperture radius of 2.2'' for the magnitude measurements and 1.5'' for the colors. The background sky is always determined in an annulus with an inner radius of 3.5'' and an outer radius of 5.0''. No aperture corrections have been applied to the extended objects, such as the star cluster candidates. Once again, these magnitudes have been photometrically calibrated to the SDSS standard system. To derive accurate positions of the clusters and to estimate properties such as ellipticity and full width at half maximum (FWHM), we have applied the Sextractor v2.5.0 \citep{Bertin1996} image classification algorithm.

\section{A New Catalog of Star Cluster Candidates  in M33}
\subsection{Cluster Search Method}

\begin{figure}
\epsscale{0.9}
\centering
\includegraphics[width=0.4\textwidth]{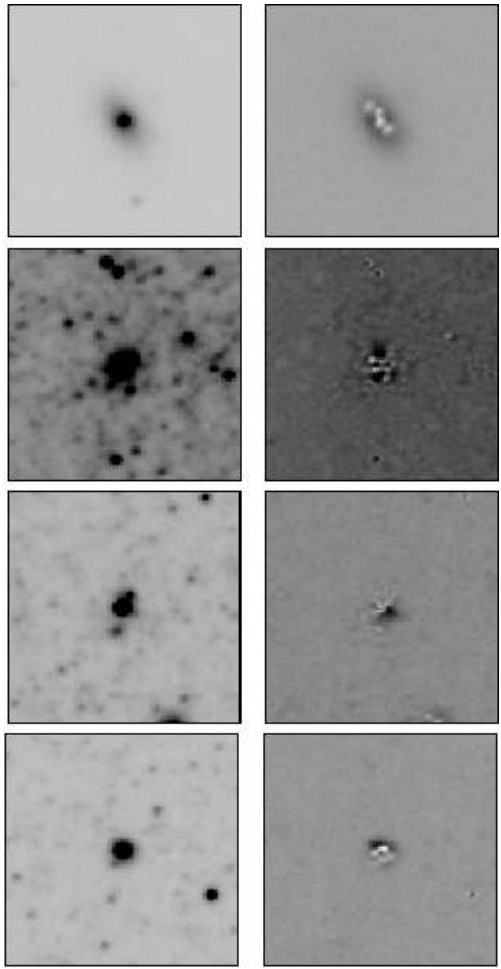}
\caption{Illustration of the search method adopted. First column corresponds to the original images while the second column corresponds to the residual images after PSF subtraction. Rows from top to bottom show: a background galaxy, a confirmed star cluster, and two new candidate clusters. Each image is shown with the same gray-scale intensity and 15'' on a side, with north up and east to the left.}
\label{mosaic}
\end{figure}

Our detection method is based on the fact that at the distance of M33, non-stellar objects are expected to be more extended than the PSF. After subtracting the stellar PSF from all of the sources in our frames, extended objects leave a doughnut-shaped appearance, as they are under-subtracted in the wings and over-subtracted in the center. We have used DAOPHOT/ALLSTAR \citep{Stetson1994} to produce residual images free of all PSF sources. We have trained our eyes to recognize the residual pattern of candidate clusters. While most background galaxies show either a spiral arm structure or an elongated pattern, the candidate stellar clusters show some level of assembly. As an illustration, Fig. \ref{mosaic} shows original and residual images of different types of extended objects. After visual inspection of the residual images as well as analysis of the original ones, this technique leaves us with a total of 2,990 extended objects: 803 candidate clusters, 1,969 galaxies and 218 unknown objects. From the total number of candidate clusters, 204 were previously identified clusters in the SM updated website and considered confirmed clusters based on HST and high-resolution ground-based imaging.  \\

The 12 ACS/HST fields examined in \cite{Sanromanetal2009} included 72 of the candidate clusters in the present catalog, where 51 turned out to be genuine star clusters. This suggests that around $\sim$ 70\% of the proposed candidates will be actual stellar clusters. However, from the 349 guaranteed clusters listed in the updated SM catalog, our catalog only recovers 204 objects implying missing objects mostly in the center of the galaxy, not surprising since the method is less effective in extremely crowded regions.\\ 

\begin{figure}
\epsscale{0.9}
\centering
\includegraphics[bb=60 407 620 926, width=0.8\textwidth]{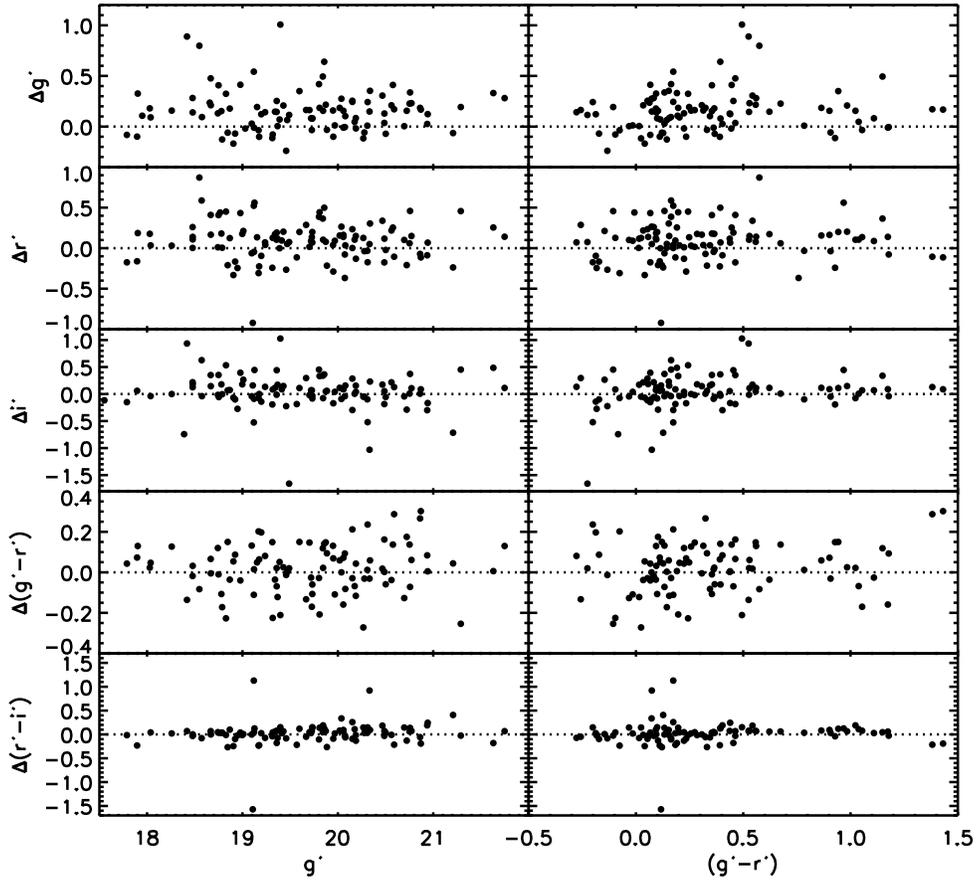}
\caption{Comparison of the integrated cluster photometry from the present study with the common objects from \cite{Zloczewskietal2008}.}
\label{photometry}
\end{figure}

Comparison with the similar study of ZKH reveals unexpected discrepancies. From the total 1,752 common objects between both catalogs, only 124 sources were classified as candidate clusters by both authors. As \cite{Sanromanetal2009} argue, the total number of 'true' cluster candidates in the ZKH catalog is not likely to be larger than $\sim$ 40\%. This suggests a systematic misidentification in the candidate object pattern or a defective PSF subtraction. Fig. \ref{photometry} shows the photometric differences between the two studies. We find a mean difference of $<$$\Delta$g$>$ = 0.15 $\pm$ 0.02, $<$$\Delta$r$>$ = 0.10 $\pm$ 0.02 and $<$$\Delta$i$>$ = 0.04 $\pm$ 0.03 while the offsets for the colors are  $<$$\Delta$(g-r)$>$ = 0.018 $\pm$ 0.012 and $<$$\Delta$(r-i)$>$ = 0.037 $\pm$ 0.030. The disagreement in the magnitude offsets disappears in the color offsets, which indicates that the photometric variation corresponds to the different adopted apertures in each study. These photometric differences are not unexpected for integrated photometry of extended objects such as star clusters \citep{Sanromanetal2009}.  

\subsection{Highly Probable Clusters}
\begin{figure}
\epsscale{0.9}
\centering
\includegraphics[width=0.49\textwidth]{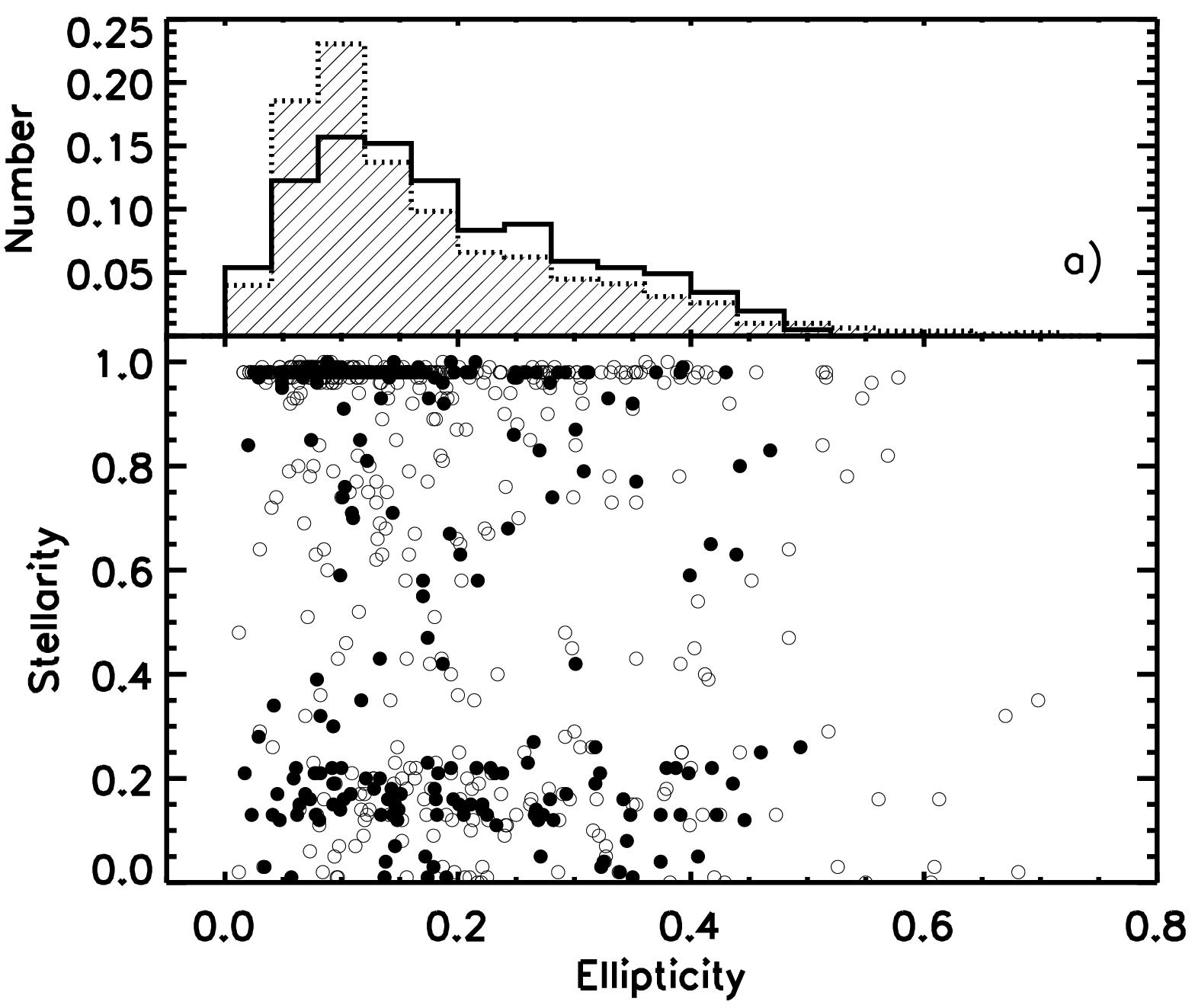}
\includegraphics[width=0.49\textwidth]{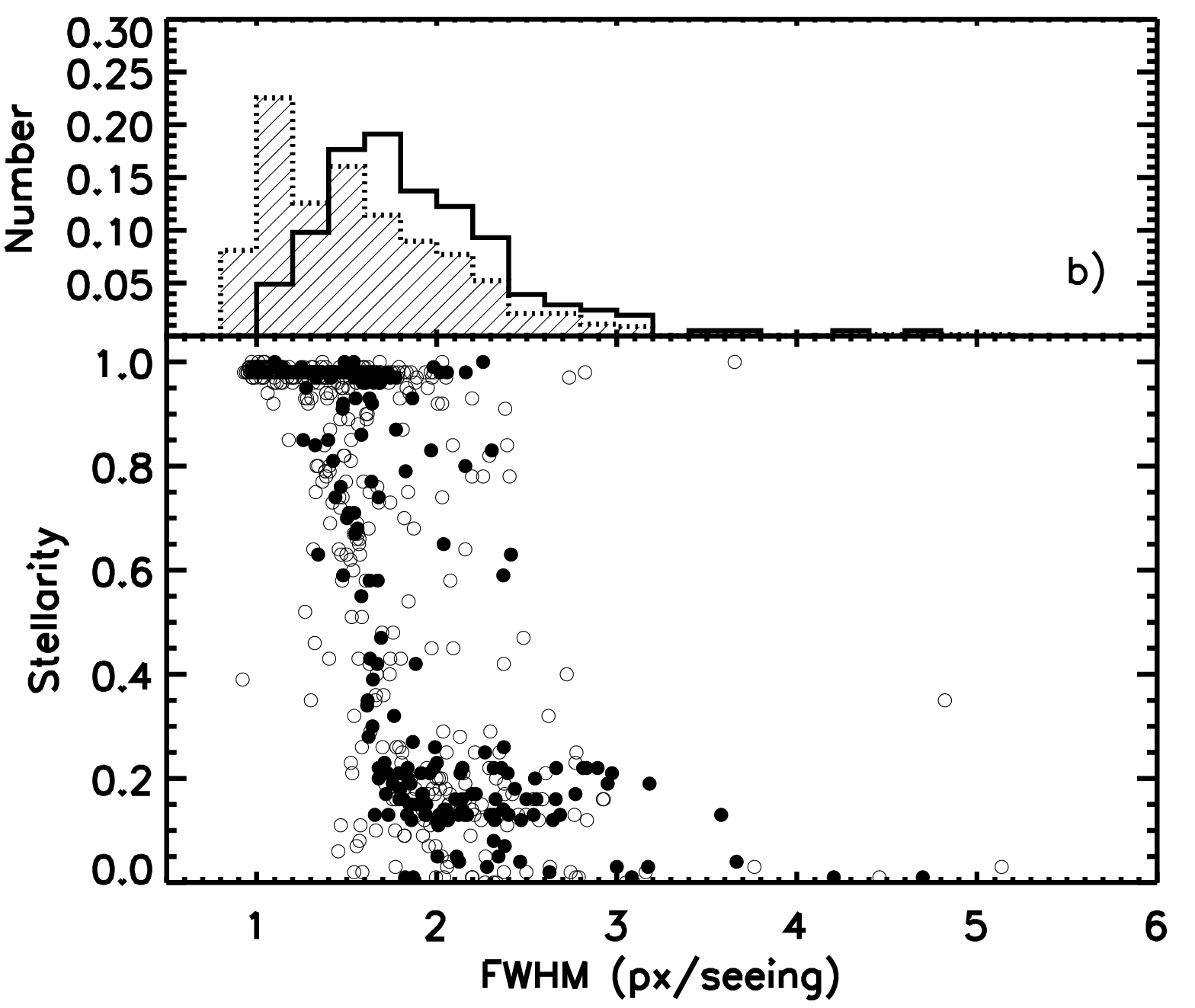}
\includegraphics[width=0.49\textwidth]{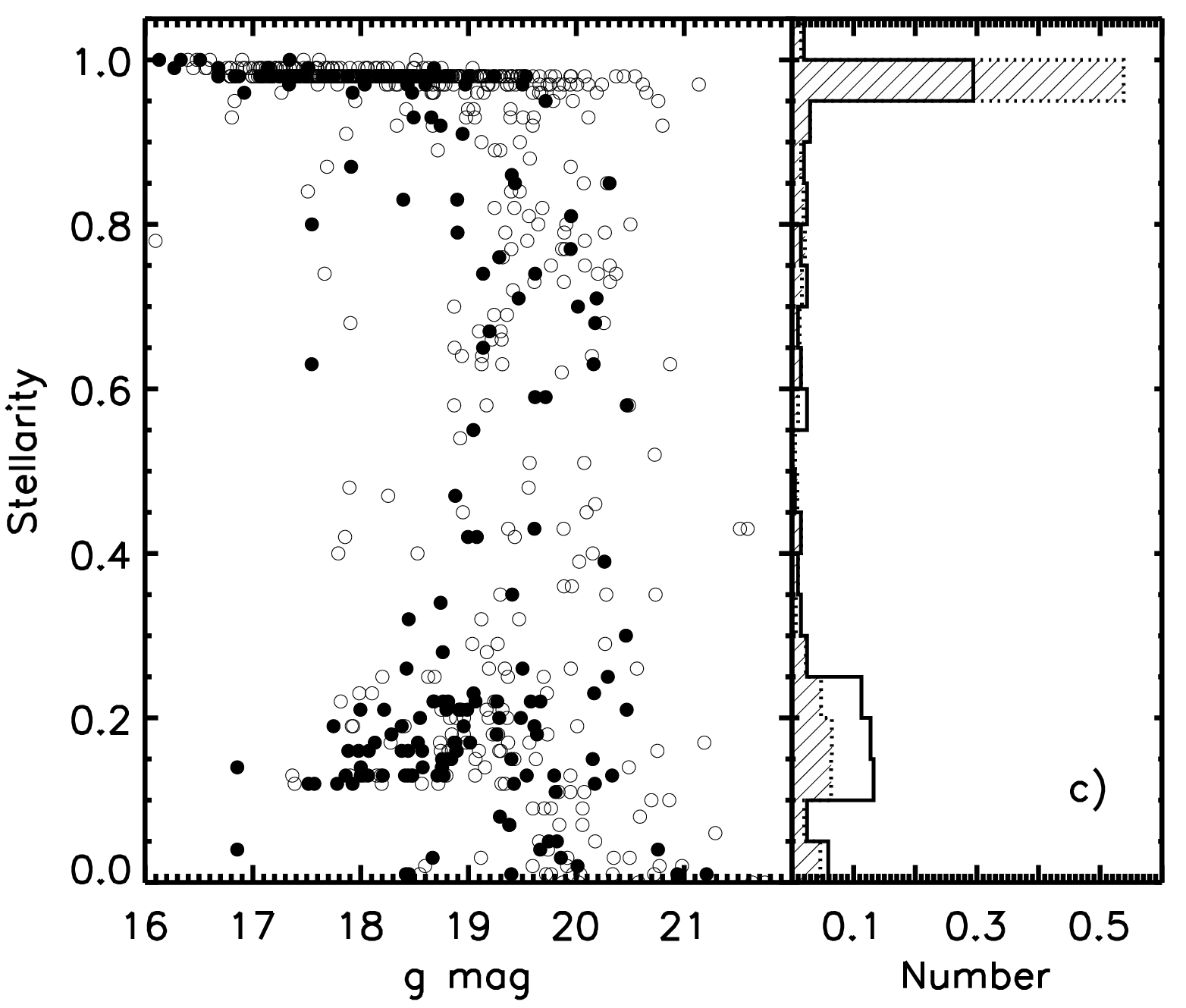}
\includegraphics[width=0.49\textwidth]{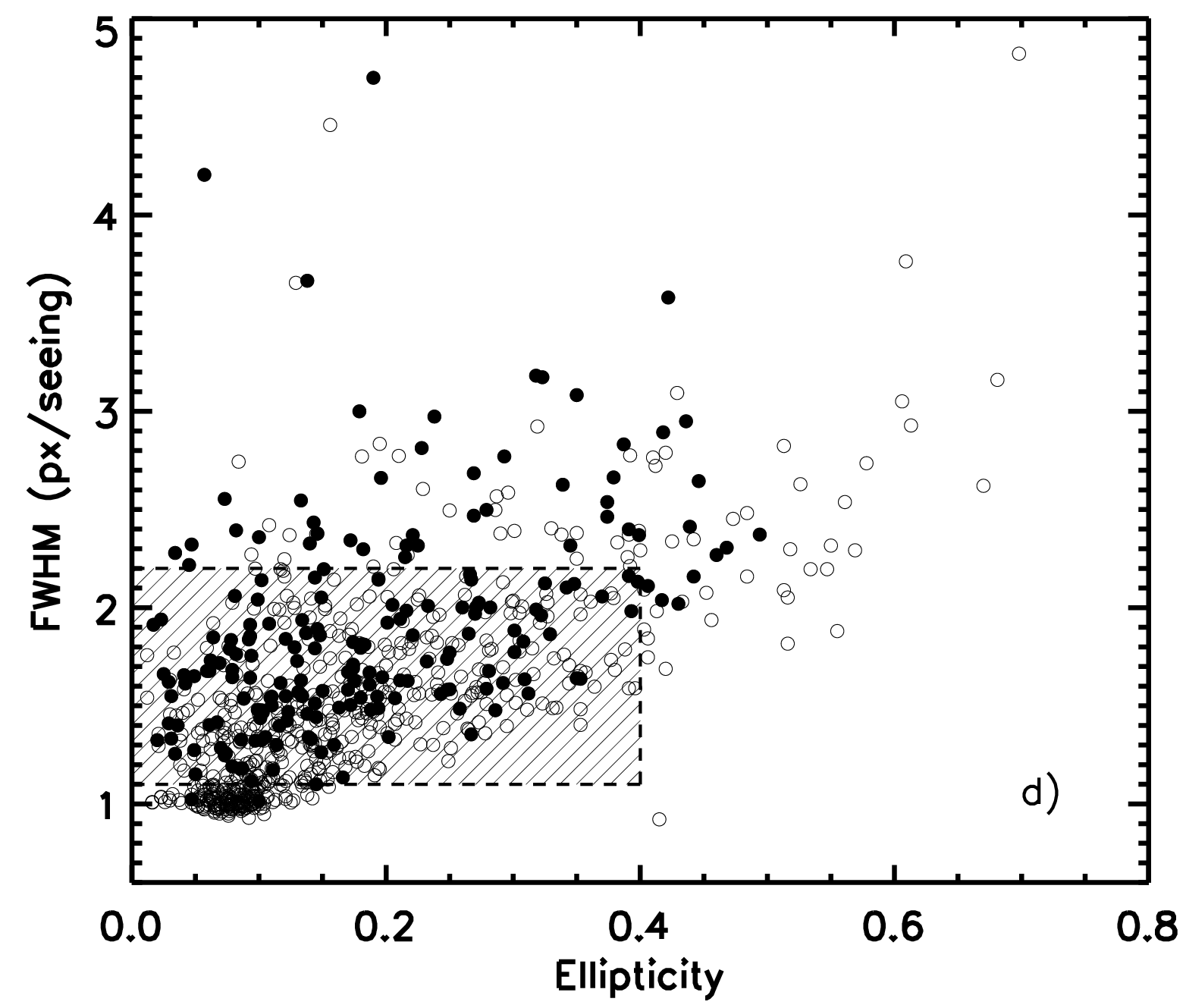}
\caption{Sextractor parameter distributions of the 599 new candidate clusters (open circles) and the 204 previously confirmed clusters in the catalog of SM (filled circles). The open histograms correspond to the distribution of confirmed clusters while the filled histograms correspond to our sample of candidate clusters. The filled area in panel d) shows the selected sample of highly probable clusters.}
\label{subsample}
\end{figure}

We have obtained stellarity, full width at half-maximum (FWHM), and ellipticity for the total sample of extended sources by applying the Sextractor software to the target images. We have compared the photometric parameters of the known M33 star clusters in our sample with the parameters of candidate objects to find a suitable criterion to select highly probable clusters. Fig. \ref{subsample} shows the distribution of the three Sextractor parameters. The stellarity parameter object classification of Sextractor allows us to examine our visual object classification with a more systematic algorithm. Based on its definition, a stellarity of 1 corresponds to a point source (star) and a stellarity of 0 to a resolved object (galaxy). Considering the pixel scale of the CCD (0.187"$pixel^{-1}$), a typical seeing of 0.7'' and the distance of M33 ($\sim$ 870 kpc), a mean cluster size of 4pc will appear in our images as a point source object of $\sim$ 0.9" implying stellarities around 1. The distribution of ellipticities peaks between e=0.05-0.2 with an extended tail reaching 0.5 in both samples. Panel b) shows the normalized FWHM assuming a mean seeing of 0.7''. The FWHM of our sample has two peaks: at FWHM $\sim$ 1.5 that agrees with the confirmed cluster distribution and another peak at FWHM $\sim$ 1.1 not associated with the confirmed cluster distribution. Fig. \ref{subsample} c) shows that the distribution of stellarity for the confirmed clusters has a strong peak at $\sim$ 1 with a weak peak $<$ 0.2. This distribution of stellarities suggests that a significant number of confirmed star clusters will be missing in the catalog if our main source of classification were the Sextractor classification algorithm. Although the stellarity is a very useful detection parameter to distinguish between point sources and non-point sources, the possibility of extended or partially resolved clusters in our images, means that the stellarity parameter must be used cautiously. \\

Based on the properties of the confirmed star clusters, we selected a sample of highly probable clusters that satisfy the following criteria: a) Ellipticity $<$ 0.4; b) 1.1 $<$ FWHM(px/ seeing) $<$ 2.2; and c) stellarity $>$ 0.6. We show in Fig. \ref{subsample} d) the correlation between FWHM and ellipticity where the filled area corresponds to the specified selection criteria. A minimum condition has been applied to the FWHM in order to avoid stellar contamination and small stellar associations. No information on color or magnitude was used for selecting the candidates. This sub-sample of highly probable clusters contains 246 objects and has been designated as class 2. Analysis of the contamination of this subsample using a similar technique as above suggests that $\sim$ 85\% of highly probable clusters will be genuine stellar clusters. To illustrate the available data, Table \ref{Table1} shows an excerpt of the complete extended source catalog where the last column corresponds to our proposed classification of the objects: 1=galaxy, 0=unknown extended object, 1=candidate star cluster, 2=highly probable cluster and 3=confirmed cluster based on the SM catalog. Based on this classification, the catalog contains 599 new candidate clusters (353 candidate clusters (class 1) and 246 highly probable clusters (class 2)). Table \ref{Table1} can be found in its entirety in the electronic edition of the Journal. The sample of highly probable clusters as well as the guaranteed clusters in SM will be used as targets for future follow-up imaging and spectroscopic observations. \\

\section{Analysis}

\subsection{Color-Magnitude Diagrams}

\begin{figure}
\epsscale{0.9}
\centering
\includegraphics[bb= 57 405 616 844,width=0.8\textwidth]{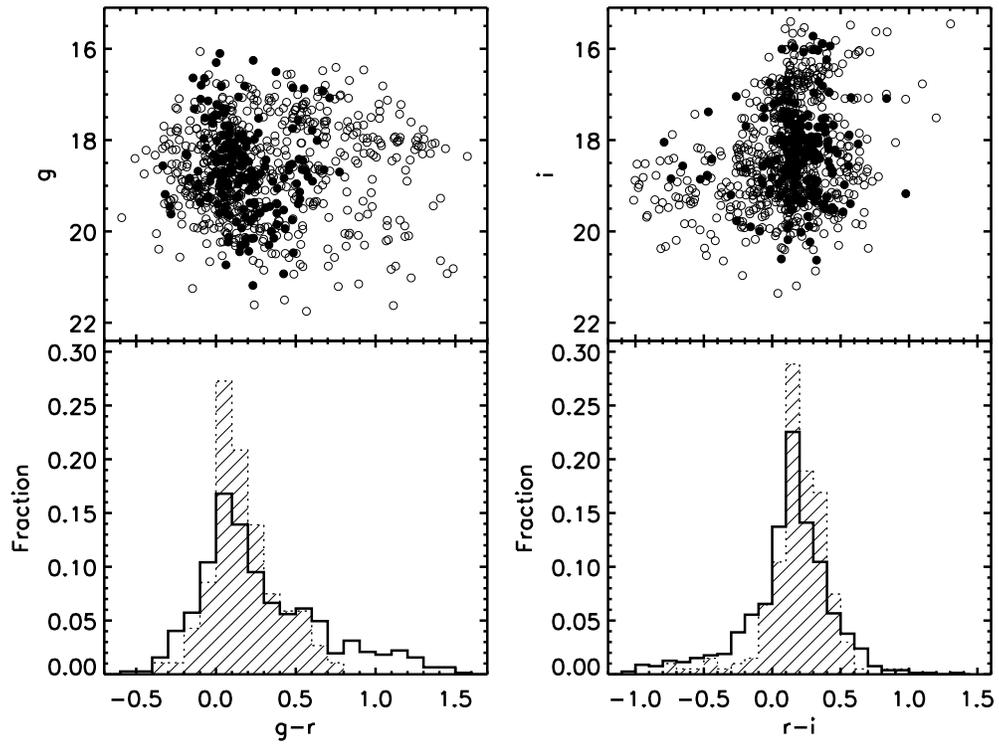}
\caption{Color-magnitude diagrams and color distributions of the new star cluster candidates (open circles/unfilled histogram) as compared with confirmed clusters from SM (filled circles/filled histogram).}
\label{color_magnitude}
\end{figure}

Intrinsic properties such as age, metallicity, and reddening govern the integrated magnitudes and colors of clusters. As described in the previous sections, we have performed aperture photometry of the candidate (class 1 and 2) and confirmed (class 3) star clusters. In addition, we have made use of the equations in \cite{Tucker2006} to transform our photometry into the SDSS \textit{ugriz} standard filters. Fig. \ref{color_magnitude} shows the color-magnitude diagrams (CMDs) and color distributions of our sample (class 1 and 2) as compared with the confirmed star clusters from the SM catalog (class 3). The magnitude distribution of our sample contains more faint star clusters than SM. The faintest clusters reach g $\sim$ 22 that  corresponds to M$_{g} \sim$ --3.0, assuming a distance modulus of $(m-M)_0$=24.69 \citep{Galletietal2004} and an average reddening correction of E(V--I)=0.06 \citep{Sarajedini2000}. The color range of our sample is significantly wider than the color range of the confirmed clusters: --0.4 $<$ (g--r) $<$ 1.5 and --1.0 $<$ (r--i) $<$ 1.0. The lower panels show a unimodal distribution with a strong peak at (g--r) $\sim$ 0.1 and (r--i) $\sim$ 0.2  having extended tails redward in (g--r) color and blueward in (r--i). 

\subsection{Color-Color Diagrams}
\begin{figure}
\epsscale{0.9}
\centering
\includegraphics[width=0.8\textwidth]{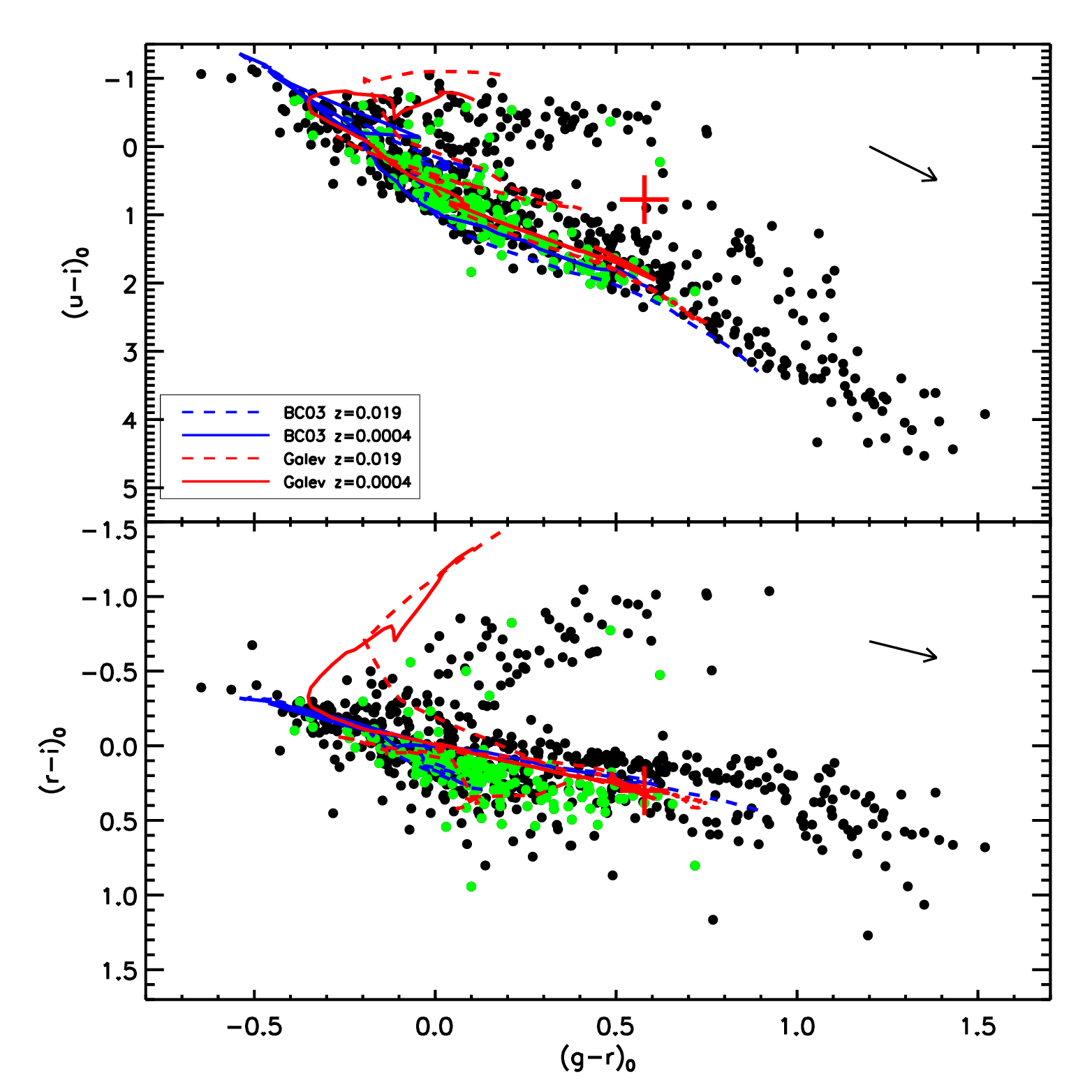}
\caption{Color-color diagram of the candidate star clusters from the present study (black circles) as compared with the confirmed clusters from SM (green circles). The solid lines correspond with the SSP models of \cite[BC03]{Bruzual2003} and \cite[Galev]{Galev} for z=0.019 and z=0.0004. All the clusters have been shifted by a line-of-sight reddening value of E(V--I) = 0.06 \citep{Sarajedini2000}. The red cross corresponds with the integrated colors of the nucleus of M33 and the arrows in the top right corners represent the direction of the reddening vector. [\textit{See the electronic edition of the Journal for a color version of this figure.}]}
\label{Galev_color_color}
\end{figure}

 Fig. \ref{Galev_color_color} shows color-color diagrams of the candidate star clusters. To compare with simple stellar populations (SSP), two different sets of models have been used: \citet[BC03]{Bruzual2003} and \citet[Galev]{Galev}. BC03 models correspond to an evolutionary track for an instantaneous burst and a Salpeter IMF while Galev models correspond to a customized set provided to us by Ralf Kotulla and the Galev team. The Galev SSP models were run assuming Geneva evolutionary tracks with a minimum age and time resolution of 0.1 Myrs until 100 Myrs, and a time-step of 1 Myrs for older ages. The models were run with a Salpeter IMF (1-120 M$_{\odot}$) for different metallicities. It is important to note that Galev models include contributions from nebular emission, considering the continuum nebular and also emission lines. All of the clusters have been shifted by a line-of-sight reddening value of E(V--I)=0.06 \citep{Sarajedini2000} and adopting an extinction relation from \cite{Cardelli1989}.\\

Comparisons between the integrated cluster colors and the predictions of stellar population models can provide age estimates that are potentially useful for studies of galaxy evolution. However, uncertainties in ages derived from multi-color photometry come not only from the photometric errors, but also from reddening corrections and uncertainties in the metal abundance of each cluster. Furthermore, the face-on view of the galaxy and the numerous spiral arms produce a broad range of reddening in M33 that can scatter the integrated colors of individual clusters. Another effect that is important in this regard is the dispersion in the integrated colors due to stochastic effects and these can vary significantly along the age sequence \citep{Girardi1995}. The adopted IMF in the SSP also contributes to uncertainties in the models. Given these points, we have not attempted to estimate ages based on the integrated photometry of the clusters. In any case, the color-color diagrams reveal a number of interesting features. \\

\begin{figure}
\epsscale{0.9}
\centering
\includegraphics[bb=139 381 466 719, width=0.8\textwidth]{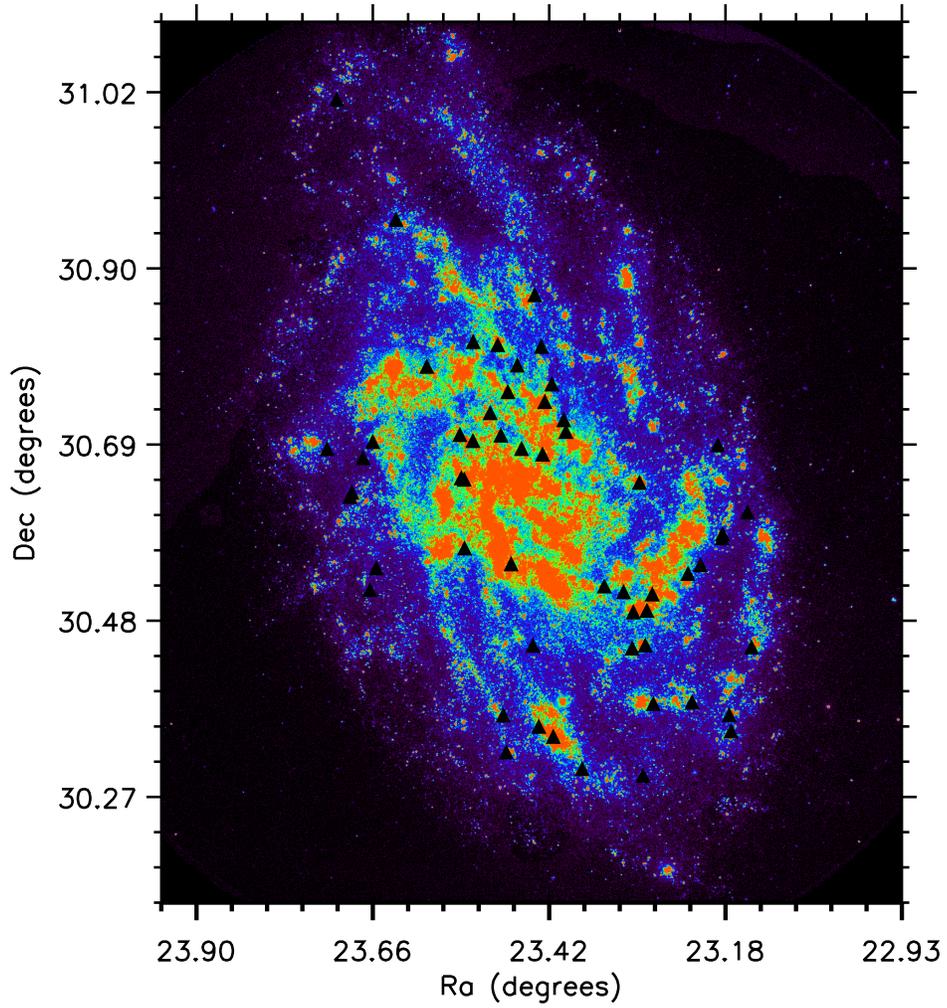}
\caption{Galex FUV image of M33. The image traces the star formation regions of the galaxy through its young stars. The black triangles correspond to the 56 star cluster candidates associated with the finger-like feature. [\textit{See the electronic edition of the Journal for a color version of this figure.}]}
\label{Galex_finger}
\end{figure}

A significant fraction of `bluish' clusters that occupy a unique location in the diagram appear in both panels of Fig. \ref{Galev_color_color} at colors (r--i) $<$ --0.2 and (u--i) $<$ 0.8. These clusters represent a finger-like feature that deviates from the expected direction of evolution. At least five confirmed clusters from SM are associated with the feature, supporting the genuine cluster nature of these objects. The age estimates for three of them shows ages $\sim$10${^7}$yrs. Based on the lower panel in Fig. \ref{Galev_color_color}, the 'finger' feature could be associated with the presence of a significant population of very young clusters ($< 10^{7}$yrs) exhibiting nebular emission. The position of many of them below the theoretical line is consistent with internal reddening shifting their colors to redder values consistent with the dusty clouds in which they are born. Fig. \ref{Galex_finger} shows the spatial distribution of these very young clusters on a Galex FUV image. A close inspection of this image suggests that all the clusters are associated with regions of star formation activity.    \\ 

The recent work of \cite{Grossi2010} analyzes multi-wavelength observations of 32 young star clusters and associations in M33. The sample was selected from catalogs of emission line objects based on their round shape and their position in regions of the galaxy that are not too crowded in the H$\alpha$ map. All of the objects have oxygen abundances of 8 $<$ 12 + log(O/H) $<$ 8.7 and have 24$\mu$m counterparts in the \textit{Spitzer/MIPS} map. Comparison of the \cite{Grossi2010} catalog and ours reveals 10 common objects, all of them associated with the previously mentioned 'finger' feature. Table \ref{Table2} provides a cross-identification of  the common objects and includes the ages, extinctions and reddenings obtained from their spectral energy distribution (SED) fitting technique. This result confirms the young age of these clusters, younger than $\sim$ 12 Myr, and their relatively high extinction, A$_{v}$, between 0.5 and 0.9. Of the 10 common objects, 2 of them have been previously confirmed as genuine star clusters.\\

Prior to this study, the M33 cluster system presented an age range of star clusters between  10 Myrs -- 10 Gyrs with the majority of clusters with ages around 100 -- 400 Myrs. Fig. \ref{Galev_color_color} reveals a wider age range of 1 Myrs -- 10 Gyrs with at least $\sim$ 50$\%$ of objects corresponding to young clusters ($<$ 100 Myrs) and $\sim$ 10$\%$ of the total corresponding to very young clusters with nebular emission (finger-like feature). In spite of the redward tail at (g--r) $>$ 0.7 that is probably caused by reddening, the diagram suggests the presence of an old population at least as old as 10 Gyr.\\   

Furthermore, the integrated colors of the nucleus of M33 (RA= 01:33:51.02; DEC= 30:39:36.68) have been plotted in the color-color diagrams of Fig. \ref{Galev_color_color} and Fig. \ref{Galev_gal}. After examining the curve of growth for different aperture diameters, we have adopted an aperture radius of 4.4'' for the magnitude measurements and a background sky of 7.5'' and 9''. The integrated light shows a blue nucleus with u=14.9, (u--i)=0.92, (u--g)=0.63 and (r--i)=0.33. 

\subsection{A Gap in the Diagram}

\begin{figure}
\epsscale{0.9}
\centering
\includegraphics[width=0.8\textwidth]{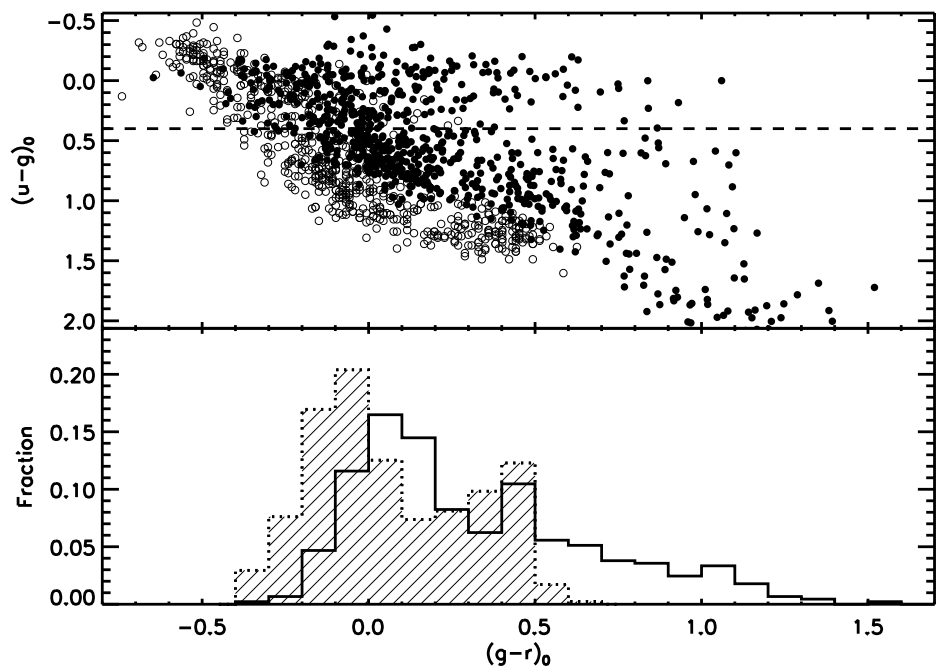}
\caption{\textit{Upper panel:} Color-color diagram of our sample (filled circles) and LMC (open circles) \citep{Bica1996} in the relevant color ranges. To avoid contamination by clusters that exhibit nebular emission, only clusters with (u--g) $>$ 0.4 (dashed line) have been considered in the color distribution. \textit{Lower panel:} Color distribution of our sample (unfilled histogram) and LMC (filled histogram). M33 gap can be detected at (g--r) $\simeq$ 0.3 and (u--g) $\simeq$ 0.8., and the LMC gap at (g--r) $\simeq$ 0.3 and (u--g) $\simeq$ 1.3.}
\label{gaps}
\end{figure}

The color-color diagrams plotted herein reveal a gap in the distribution of star clusters centered at (g--r) $\simeq$ 0.3 and (u--g) $\simeq$ 0.8. A similar anomaly was discovered by \cite{Bica1991} among LMC star clusters in the (U--B) vs. (B--V) diagram. The LMC gap was noticed at (U--B) $\simeq$ 0.19 and (B--V) $\simeq$ 0.47 with an approximate width of 0.1 mag in both colors. The upper panel in Fig. \ref{gaps} shows a small region of the LMC and M33 color-color diagrams for better visualization of the gaps. The LMC data are from \cite{Bica1996} and have been converted to the SDSS \textit{ugriz} system using relations published by \cite{Jester2005}. Constant reddening values of E(V--I)=0.06 \citep{Sarajedini2000} and E(B--V)=0.1 \citep{Alves2002} have been adopted for M33 and LMC, respectively. Although the M33 gap appears bluer in the diagram, both gaps correspond to a similar range in age. The lower panel in Fig. \ref{gaps} shows the color distribution of the M33 and LMC clusters. To avoid contamination by clusters in the finger-like feature, which corresponds to a different age range, only clusters with (u--g) $>$ 0.4 have been considered in the construction of the color distribution. The gap in M33 looks smaller and slightly redder than the gap in the LMC. The offset could be a consequence of the reddening correction we have applied which does not account for dust internal to each galaxy. 

Several authors \citep{Renzini1986, Sweigart1990} have interpreted the LMC gap as being produced by the red giant branch phase transition. This transition would be produced by stars at the helium flash stage and would fit theoretical predictions. 
\cite{Girardi1995} disagrees with this interpretation and argues that the lack of clusters in this region is determined by the natural dispersion of the colors. The stochastic effects on the mass distribution of stars could produce the dispersion in the colors so no additional peculiarities would be needed in the stellar models in order to reproduce this feature of the diagram. However, the discovery of the gap among M33 clusters supports the presence of an evolutionary effect at that particular age as the origin of both gaps.

\subsection{Additional Comparisons With Other Galaxies}

\begin{figure}
\epsscale{0.9}
\centering
\includegraphics[width=0.8\textwidth]{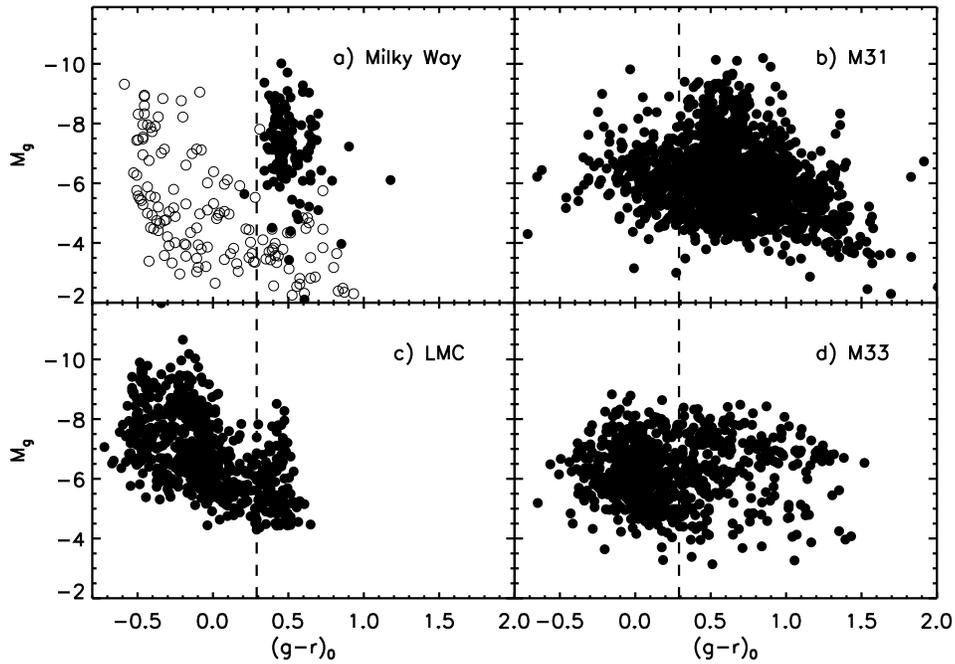}
\caption{Color-magnitude diagrams of star clusters in different galaxies of the Local Group: a) The Milky Way: open clusters (open circles) \citep{Lata2002} and globular clusters (filled circles) \citep{Harris1996}; b) M31 \citep{Peacock2009} c) LMC \citep{Bica1996} and c) M33 confirmed and candidate star clusters from this study. The dashed lines represent the division of Galactic globular clusters at (B--V)$_{0}$ = 0.5.}
\label{cmd_comparison}
\end{figure}

\begin{figure}
\epsscale{0.9}
\centering
\includegraphics[width=0.8\textwidth]{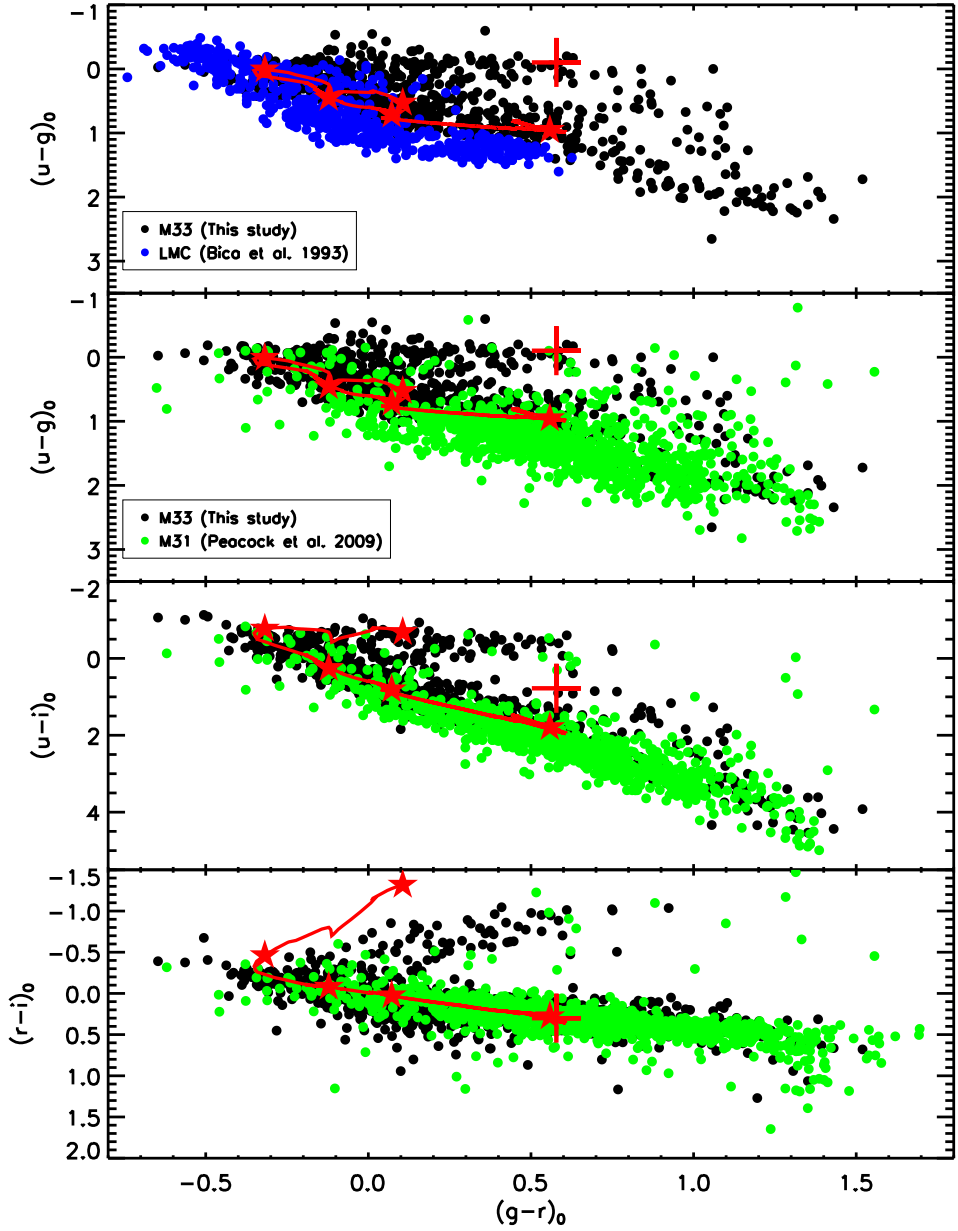}
\caption{Color-color diagrams of the candidate star clusters from the present study (black circles) as compared with the candidate and confirmed clusters in M31 (green circles) \citep{Peacock2009} and confirmed clusters in the LMC (blue circles)\citep{Bica1996}. The solid line corresponds to the SSP models of \cite{Galev} with a metallicity of z=0.0004. The star symbols correspond to ages of 10$^{6}$, 10$^{7}$, 10$^{8}$, 10$^{9}$ and 10$^{10}$ yrs and the red cross corresponds to the integrated colors of the nucleus of M33. [\textit{See the electronic edition of the Journal for a color version of this figure.}] }
\label{Galev_gal}
\end{figure}

Fig. \ref{cmd_comparison} presents the color-magnitude diagrams of the cluster system of our sample, the Milky Way (MW), M31 and the Large Magellanic Cloud (LMC). The MW data belong to \cite{Lata2002} (open clusters) and \cite{Harris1996} (globular clusters). The M31 data correspond to the candidate and confirmed clusters in \cite{Peacock2009}.
Data for the LMC cluster system from \cite{Bica1996} have also been plotted. In order to compare the different cluster systems, we have plotted absolute magnitudes and reddening corrected colors. The MW absolute magnitudes are taken directly from the above-mentioned catalogs assuming a specific distance modulus and reddening for each cluster. We have adopted a LMC distance modulus of 18.50 \citep{Freedman2001} and 24.36 for M31 \citep{Vilardell2010}.
 Constant reddening values of E(V--I)=0.06 \citep{Sarajedini2000}, E(V--I)=0.1 \citep{Durrell2001} and E(B--V)=0.1 \citep{Alves2002} have been adopted for M33, M31 and the LMC, respectively. If needed, we have used the Jester et al (2005) transformations to convert absolute magnitudes into the g-band. The dashed lines represent the division of Galactic globular clusters at (B--V)$_{0}$ = 0.5.\\      

No distinct cluster subpopulations can be identified within the M33 cluster system like in the LMC or MW. However, the integrated colors of the very young clusters are not necessarily a reflection of their ages because they could be affected by nebular emission. Considering the significant number of this type of object in our sample, the color distribution could be distorted and appear unimodal when in fact it is not. When the nebular emission clusters are removed from the analysis (see Fig. \ref{gaps}), the color-magnitude diagram shows a possible bimodality. The M33 and LMC systems are dominated by blue clusters, (B--V)$_{0}$ $<$ 0.5, in contrast with the redder M31 system. However, while the red cluster subpopulation of LMC occupies a very narrow (g--r)$_{0}$ region, M33 red clusters populate a significantly wider color range more similar to M31 red clusters.\\ 

When comparing the absolute magnitudes of the cluster systems, we see that the brightest clusters in the MW and LMC reach luminosities of M$_{g}$ $\sim$ --9.5; however, the brightest clusters in M33 correspond to M$_{g}$ $\sim$ --8, more than one magnitude fainter. This effect could be explained by the relation between the star formation rate (SFR) of a galaxy and the maximum mass/luminosity of its star clusters \citep{Larsen2002}. The empirical relation suggests that galaxies with high SFRs form proportionally more clusters, and as a consequence, the cluster mass function reaches higher masses. Assuming a SFR of $\sim$ 0.45 M$_{\odot}$ yrs$^{-1}$, the cluster system of M33 would fit onto this relation reasonably well (see Fig. 1 \cite{Bastian2008}). With the slightly higher SFR for the MW, LMC and M31 (eg. \cite{Kang2009}, \cite{Robitaille2010}) these systems will produce brighter clusters. In addition, environmental variations, such as the mass/luminosity of the galaxy, can play a role in the color-magnitude diagram of a cluster system \citep{Mieske2010}.\\

Fig. \ref{Galev_gal} presents the color-color diagrams of the M33 cluster system using our sample, M31 and LMC. The sources of the data are the same as those given above. As a reference, SSP models from the Galev team \citep{Galev} with a metallicity of z=0.0004 have been overplotted. To identify different time periods, the star symbols correspond to 10$^{6}$, 10$^{7}$, 10$^{8}$, 10$^{9}$ and 10$^{10}$ yrs. The same constant reddening value has been adopted for each sample as in the previous figure.\\

The wide color range of the M33 clusters, --0.4 $<$ (g--r) $<$ 1.5, overlaps entirely with the young-intermediate age system of the LMC and with the older M31 system. The broad range of colors implies a large range of ages, suggesting a prolonged epoch of formation. Based on this evidence, the majority of the clusters will be young-intermediate age objects although we would expect clusters older than 10 Gyrs. The diagram also shows that a small group of M31 clusters occupies the unique area of the `finger' feature, however the region seems to be significantly more populated in M33 than in these two galaxies.\\

When comparing M33 with similar morphological type galaxies such as NGC 300 or M101, M33 seems to posses a unique very young star cluster population. The color distribution of candidate clusters in M101 is similar to M33 candidate clusters but no evidence of very young clusters with nebular emission has been found \citep{Barmby2006, Chandar2004}. Although NGC 300 is nearly a twin galaxy of M33 in terms of Hubble type and mass, there are several differences between them \citep{Gogarten2010}. NGC 300 appears to have globular clusters similar to those of the Milky Way \citep{Nantais2010} and a metallicity gradient consistent with stars formed prior to 6 Gyrs ago \citep{Gogarten2010}. Environmental factors may play a key role in the star formation history of M33, as NGC 300 is isolated from other galaxies while M33 appears to be interacting with M31 \citep{McConnachie2009, Putman2009}. \cite{McConnachie2009} propose a plausible M31-M33 interaction model that reproduces with good agreement the observed distances, angular positions and radial velocities of these galaxies as well as the well-known HI warp in M33. In this simulation, M33 starts its orbit around M31 about 3.4 Gyrs ago reaching pericenter (r $\sim$ 56 kpc) around 2.6 Gyrs ago. After it passes apocenter (r $\sim$ 264 kpc) about 900 Myrs ago, M33 would be approaching M31. This close encounter could have triggered an epoch of star formation in M33. The significant population of very young clusters with nebular emission and their association with star formation regions are evidence in support of recent star formation activity in M33. \\

Many studies have shown that interacting/merger environments form large populations of clusters (e.g.\cite{Whitmore1995}; \cite{Gallagher2001}), especially very young clusters. We would expect to see very young clusters still embedded in their dust cocoons in these disturbed systems. Yet, their color-color diagrams do not exhibit as prominent a finger-like feature due to nebular emission around very young clusters as compared with M33 (see Antenna \citep{Whitmore1995}; Stephan's quintet \citep{Gallagher2001}). In the unusual environment of Hickson compact group 31 (HGC 31), \citet{Gallagher2010} found a large population of $<$ 10 Myr star clusters with strong nebular emission, similar to the one found in the present study. The main galaxies that make up HGC 31 are disrupted under the  presence of strong gravitational interactions and show tidal structures. The star cluster candidates with nebular emission appear throughout HCG31, specifically concentrated in the interaction regions. The existence of these very young star clusters seems to be the consequence of active recent and ongoing star formation in HGC 31. The similarities between the M33 cluster system and that of HGC 31, which is a strongly interacting environment, support two important assertions. First, the finger-like feature is a genuine characteristic and not an artificial effect due to the contamination of our cluster sample. Second, the past interactions between M33 and M31 have likely had significant impact on the properties of the M33 cluster system, especially the youngest clusters.

\subsection{Spatial Distribution}

In order to analyze the spatial distribution of clusters with different ages, we have divided our sample into two groups based on comparisons with SSP models. Based on BC03 models, clusters with (r--i)$_{0}$ $\sim$ 0 and (g--r)$_{0}$ $\sim$ 0.1 have ages of $\sim$ 10$^{8}$yrs.  We are going to consider clusters with (r--i)$_{0}$ $ >$ 0 and (g--r)$_{0}$ $>$ 0.1 as red or old clusters. The remaining clusters we categorize as blue or young objects. This partition minimizes the contamination of the old (red) clusters by the young clusters in the finger-like feature which exhibit integrated colors that are redder than expected. In order to place the cluster density distribution in the context of the field stars, we have made use of the \citet{Hartman2006} star catalog constructed over the same MegaCam/CFHT images used in the present study.\\

\begin{figure}
\epsscale{0.9}
\centering
\includegraphics[width=0.8\textwidth]{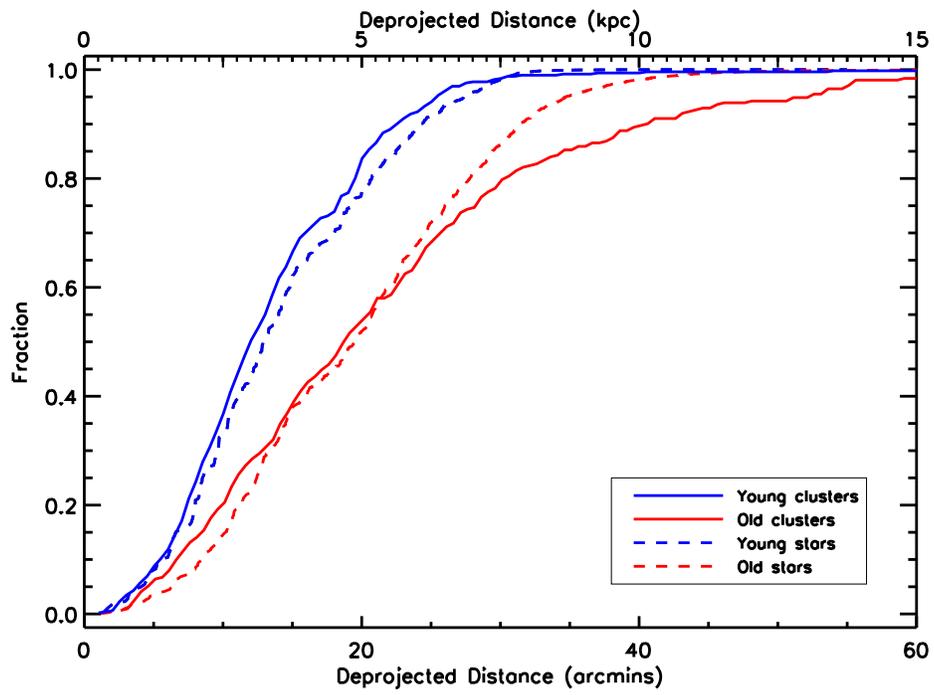}
\caption{Cumulative radial distribution for the star clusters (this study) and field stars \citep{Hartman2006} in M33. [\textit{See the electronic edition of the Journal for a color version of this figure.}]}
\label{cumulative}
\end{figure}

Fig. \ref{cumulative} shows the cumulative radial distributions of the young and old cluster populations as compared with the blue (young) and red (old) field star populations. Blue clusters follow a spatial distribution similar to the blue field stars.  The distribution also suggests that younger (bluer) clusters are more centrally concentrated as compared with older (redder) clusters. The red clusters are more dispersed in a wider region than the bluer ones, indicating that the majority of the red (old) clusters likely belong to the halo while the bluer (younger) clusters generally belong to the disk of M33. Analysis of these distributions using a Kolmogorov-Smirnov (K--S) test shows that there is a greater than 99.9$\%$ chance that the old cluster population is significantly different than the young cluster population.\\ 

\begin{figure}
\epsscale{0.9}
\centering
\includegraphics[width=0.8\textwidth]{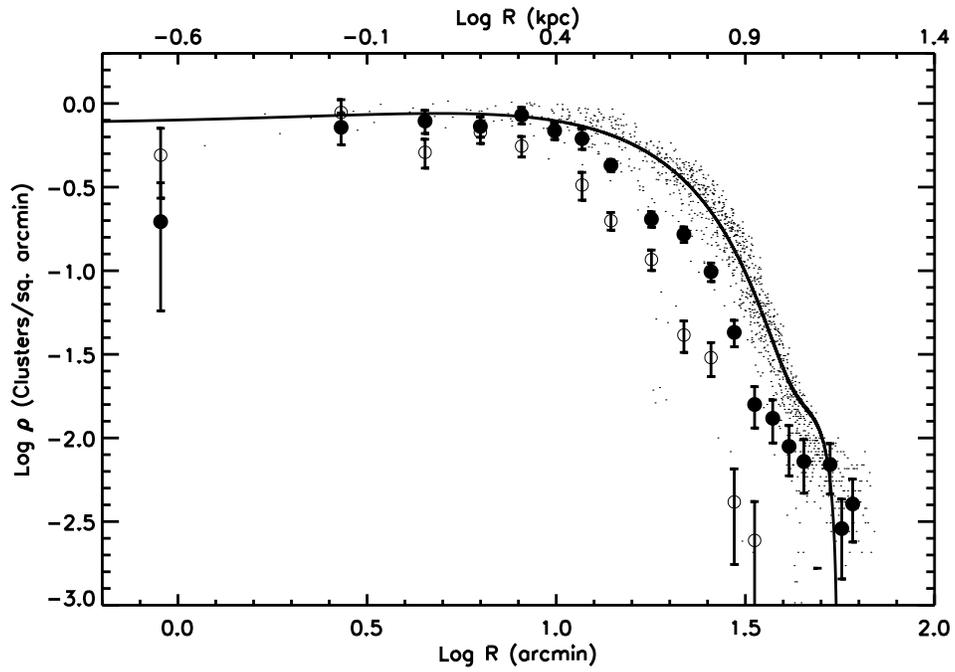}
\caption{Radial density distribution of our cluster sample (filled circles) as compared with the field stars (small dots) from \cite{Hartman2006} versus deprojected radius. The solid line represents the best polynomial fit of the field star radial density. As a comparison with previous catalogs, open circles correspond to the confirmed clusters in SM.}
\label{radial_distribution}
\end{figure}

Fig. \ref{radial_distribution} shows the radial density distribution of our entire cluster sample. The filled circles show the cluster density profile versus deprojected radius, assuming our adopted distance modulus of $(m-M)_0$=24.69, while the open circles show the confirmed clusters from SM for comparison. The small dots correspond with the radial density distribution of the field stars where the solid line represents the best polynomial fit. The star density distribution has been scaled to match the cluster density in the region between R=0.6 -- 2 kpc where both distributions are likely to have similar completeness levels.\\

Inside $\sim$ 0.8 kpc, the cluster profile presents a decrease in density, suggesting some level of incompleteness. The cluster profile outside of $\sim$ 2.5 kpc could be reproduced by a power-law where the most distant clusters are located at $\sim$ 15 kpc (63 arcmin) from the center of the galaxy. Confirmed clusters from SM show that the M33 cluster system seems to be more centrally concentrated than the field stars, however no other galaxy has been found with this characteristic \citep{Forbes1996}. Our radial profiles, shown in Fig. \ref{radial_distribution}, have reduced the discrepancy between the clusters and field stars but the former are still more centrally concentrated than the latter as shown by the cumulative distributions in Fig. \ref{cumulative}. The pronounced decrease at $\sim$ 15 kpc in the cluster and field-star density distributions suggest that this distance may represent the outer edge of both distributions.
For a given radial bin in the outer region of the galaxy, the density of clusters is significantly lower than the density of stars. We note the possibility that the cluster and stellar samples may have different completeness properties. In order to minimize the potential impact of incompleteness, we restrict the comparison of these samples to the region outside $\sim$ 0.8 kpc from the center of the galaxy. If the incompleteness of our sample is the reason for the differences between the cluster and field star radial profiles, then we would expect a random bias or perhaps a larger incompleteness toward the center of the galaxy. However, the analysis shows a significantly lower density of clusters between R = 3 -- 9 kpc than in the inner region between R=0.8 -- 3 kpc. 

The ratio of stars to clusters is determined not only by the formation  processes but also by the destruction processes. If we assume that the formation of star clusters and the formation of stars in a galaxy are correlated, then Fig. \ref{radial_distribution} suggests that the cluster system in M33 has suffered from destruction or depletion of clusters at specific radii. Tidal interactions when passing through the disk or near massive objects such as giant molecular clouds could produce tidal shocks that lead to the ultimate destruction of a cluster (e.g. \cite{Gieles2006}; \cite{Lamers2005}). Analysis of the dynamical evolution of these clusters is needed to reveal the level of influence of these interactions in the disruption process. Other environment effects such as interactions between M33 and M31 can also play a role in the depletion or disruption of clusters at preferred galactocentric distances. \cite{Huxor2009} discovered the presence of four new outlying star clusters in M33 which have large projected radii of 38 -- 113 arcmin (9.6 -- 28.5 kpc). Based on the asymmetry in the distribution of these outer clusters, they suggest the possibility that interactions with M31 may have dramatically affected the population of M33 star clusters. Regardless of the source of this anomaly, we would need an additional $\sim$ 350 clusters between R=3 -- 9 kpc in order to match the stellar density in the same region of the galaxy.

If we rescale the density of stars to match the cluster density in the outer region, a notable excess of clusters occurs at R $<$ 4 kpc. This scenario is highly unlikely since dynamical destruction processes are more effective near the central region of a galaxy. The short lifetime of such a young sample of clusters also makes the cluster migration scenario implausible. No case has been found in which the cluster density exceeds the star density in the inner region of a galaxy. Future follow-ups of this sample will test the validity of the depletion phenomenon that could have widespread repercussions for our understanding of M33's formation and evolution.

\section{Summary}
We present a wide-field photometric survey of M33 extended objects using CFHT/ MegaCam images. The resultant catalog contains 2,990 extended sources, including 599 new candidate stellar clusters and 204 previously identified clusters. We have investigated the photometric properties of the cluster sample, performing  \textit{ugriz} integrated photometry and using their morphological parameters. Based on the properties of confirmed star clusters, we select a sub-sample of 246 highly probable objects. Analysis of multicolor photometry of the candidate clusters reveals a wide range of colors including a  finger-like feature in the color-color diagrams that deviates from the expected direction of evolution. Color distributions of the cluster sample reveal a unimodal distribution. A comparison of the radial density distribution for the field stars and our cluster sample suggests that the M33 cluster system suffers from a depletion of clusters at all radii. Color-color diagrams also reveal a gap in the distribution of star clusters similar to the gap detected among LMC clusters.

\acknowledgments
We thank Ralf Kotulla for generously provide us with the customized Galev models used in this work. We also appreciate the useful comments of Rupali Chandar. We are grateful for support from the United States National Science Foundation via grant number AST-0707277. 


\begin{thebibliography}{53}
\expandafter\ifx\csname natexlab\endcsname\relax\def\natexlab#1{#1}\fi

\bibitem[{{Alves} {et~al.}(2002){Alves}, {Rejkuba}, {Minniti}, \&
  {Cook}}]{Alves2002}
{Alves}, D.~R., {Rejkuba}, M., {Minniti}, D., \& {Cook}, K.~H. 2002, \apjl,
  573, L51

\bibitem[{{Barmby} {et~al.}(2006){Barmby}, {Kuntz}, {Huchra}, \&
  {Brodie}}]{Barmby2006}
{Barmby}, P., {Kuntz}, K.~D., {Huchra}, J.~P., \& {Brodie}, J.~P. 2006, \aj,
  132, 883

\bibitem[{{Bastian}(2008)}]{Bastian2008}
{Bastian}, N. 2008, \mnras, 390, 759

\bibitem[{{Bertin} \& {Arnouts}(1996)}]{Bertin1996}
{Bertin}, E. \& {Arnouts}, S. 1996, \aaps, 117, 393

\bibitem[{{Bertin} {et~al.}(2002){Bertin}, {Mellier}, {Radovich}, {Missonnier},
  {Didelon}, \& {Morin}}]{Bertin2002}
{Bertin}, E., {Mellier}, Y., {Radovich}, M., {Missonnier}, G., {Didelon}, P.,
  \& {Morin}, B. 2002, in Astronomical Society of the Pacific Conference
  Series, Vol. 281, Astronomical Data Analysis Software and Systems XI, ed.
  {D.~A.~Bohlender, D.~Durand, \& T.~H.~Handley}, 228--+

\bibitem[{{Bica} {et~al.}(1996){Bica}, {Claria}, {Dottori}, {Santos}, \&
  {Piatti}}]{Bica1996}
{Bica}, E., {Claria}, J.~J., {Dottori}, H., {Santos}, Jr., J.~F.~C., \&
  {Piatti}, A.~E. 1996, \apjs, 102, 57

\bibitem[{{Bica} {et~al.}(1991){Bica}, {Dottori}, {Santos}, {Claria}, \&
  {Piatti}}]{Bica1991}
{Bica}, E., {Dottori}, H., {Santos}, Jr., J.~F.~C., {Claria}, J.~J., \&
  {Piatti}, A. 1991, \apjl, 381, L51

\bibitem[{{Bruzual} \& {Charlot}(2003)}]{Bruzual2003}
{Bruzual}, G. \& {Charlot}, S. 2003, \mnras, 344, 1000

\bibitem[{{Cardelli} {et~al.}(1989){Cardelli}, {Clayton}, \&
  {Mathis}}]{Cardelli1989}
{Cardelli}, J.~A., {Clayton}, G.~C., \& {Mathis}, J.~S. 1989, \apj, 345, 245

\bibitem[{{Chandar} {et~al.}(1999){Chandar}, {Bianchi}, \& {Ford}}]{CBF1999}
{Chandar}, R., {Bianchi}, L., \& {Ford}, H.~C. 1999, \apjs, 122, 431

\bibitem[{{Chandar} {et~al.}(2001){Chandar}, {Bianchi}, \& {Ford}}]{CBF2001}
---. 2001, \aap, 366, 498

\bibitem[{{Chandar} {et~al.}(2004){Chandar}, {Whitmore}, \&
  {Lee}}]{Chandar2004}
{Chandar}, R., {Whitmore}, B., \& {Lee}, M.~G. 2004, \apj, 611, 220

\bibitem[{{Clem} {et~al.}(2008){Clem}, {Vanden Berg}, \& {Stetson}}]{Clem2008}
{Clem}, J.~L., {Vanden Berg}, D.~A., \& {Stetson}, P.~B. 2008, \aj, 135, 682

\bibitem[{{Durrell} {et~al.}(2001){Durrell}, {Harris}, \&
  {Pritchet}}]{Durrell2001}
{Durrell}, P.~R., {Harris}, W.~E., \& {Pritchet}, C.~J. 2001, \aj, 121, 2557

\bibitem[{{Forbes} {et~al.}(1996){Forbes}, {Franx}, {Illingworth}, \&
  {Carollo}}]{Forbes1996}
{Forbes}, D.~A., {Franx}, M., {Illingworth}, G.~D., \& {Carollo}, C.~M. 1996,
  \apj, 467, 126

\bibitem[{{Freedman} {et~al.}(2001){Freedman}, {Madore}, {Gibson}, {Ferrarese},
  {Kelson}, {Sakai}, {Mould}, {Kennicutt}, {Ford}, {Graham}, {Huchra},
  {Hughes}, {Illingworth}, {Macri}, \& {Stetson}}]{Freedman2001}
{Freedman}, W.~L., {Madore}, B.~F., {Gibson}, B.~K., {Ferrarese}, L., {Kelson},
  D.~D., {Sakai}, S., {Mould}, J.~R., {Kennicutt}, Jr., R.~C., {Ford}, H.~C.,
  {Graham}, J.~A., {Huchra}, J.~P., {Hughes}, S.~M.~G., {Illingworth}, G.~D.,
  {Macri}, L.~M., \& {Stetson}, P.~B. 2001, \apj, 553, 47

\bibitem[{{Gallagher} {et~al.}(2001){Gallagher}, {Charlton}, {Hunsberger},
  {Zaritsky}, \& {Whitmore}}]{Gallagher2001}
{Gallagher}, S.~C., {Charlton}, J.~C., {Hunsberger}, S.~D., {Zaritsky}, D., \&
  {Whitmore}, B.~C. 2001, \aj, 122, 163

\bibitem[{{Gallagher} {et~al.}(2010){Gallagher}, {Durrell}, {Elmegreen},
  {Chandar}, {English}, {Charlton}, {Gronwall}, {Young}, {Tzanavaris},
  {Johnson}, {Mendes de Oliveira}, {Whitmore}, {Hornschemeier}, {Maybhate}, \&
  {Zabludoff}}]{Gallagher2010}
{Gallagher}, S.~C., {Durrell}, P.~R., {Elmegreen}, D.~M., {Chandar}, R.,
  {English}, J., {Charlton}, J.~C., {Gronwall}, C., {Young}, J., {Tzanavaris},
  P., {Johnson}, K.~E., {Mendes de Oliveira}, C., {Whitmore}, B.,
  {Hornschemeier}, A.~E., {Maybhate}, A., \& {Zabludoff}, A. 2010, \aj, 139,
  545

\bibitem[{{Galleti} {et~al.}(2004){Galleti}, {Bellazzini}, \&
  {Ferraro}}]{Galletietal2004}
{Galleti}, S., {Bellazzini}, M., \& {Ferraro}, F.~R. 2004, \aap, 423, 925

\bibitem[{{Gieles} {et~al.}(2006){Gieles}, {Portegies Zwart}, {Baumgardt},
  {Athanassoula}, {Lamers}, {Sipior}, \& {Leenaarts}}]{Gieles2006}
{Gieles}, M., {Portegies Zwart}, S.~F., {Baumgardt}, H., {Athanassoula}, E.,
  {Lamers}, H.~J.~G.~L.~M., {Sipior}, M., \& {Leenaarts}, J. 2006, \mnras, 371,
  793

\bibitem[{{Girardi} {et~al.}(1995){Girardi}, {Chiosi}, {Bertelli}, \&
  {Bressan}}]{Girardi1995}
{Girardi}, L., {Chiosi}, C., {Bertelli}, G., \& {Bressan}, A. 1995, \aap, 298,
  87

\bibitem[{{Gogarten} {et~al.}(2010){Gogarten}, {Dalcanton}, {Williams}, {Ro{\v
  s}kar}, {Holtzman}, {Seth}, {Dolphin}, {Weisz}, {Cole}, {Debattista},
  {Gilbert}, {Olsen}, {Skillman}, {de Jong}, {Karachentsev}, \&
  {Quinn}}]{Gogarten2010}
{Gogarten}, S.~M., {Dalcanton}, J.~J., {Williams}, B.~F., {Ro{\v s}kar}, R.,
  {Holtzman}, J., {Seth}, A.~C., {Dolphin}, A., {Weisz}, D., {Cole}, A.,
  {Debattista}, V.~P., {Gilbert}, K.~M., {Olsen}, K., {Skillman}, E., {de
  Jong}, R.~S., {Karachentsev}, I.~D., \& {Quinn}, T.~R. 2010, \apj, 712, 858

\bibitem[{{Grossi} {et~al.}(2010){Grossi}, {Corbelli}, {Giovanardi}, \&
  {Magrini}}]{Grossi2010}
{Grossi}, M., {Corbelli}, E., {Giovanardi}, C., \& {Magrini}, L. 2010, ArXiv
  e-prints

\bibitem[{{Harris}(1996)}]{Harris1996}
{Harris}, W.~E. 1996, \aj, 112, 1487

\bibitem[{{Hartman} {et~al.}(2006){Hartman}, {Bersier}, {Stanek}, {Beaulieu},
  {Kaluzny}, {Marquette}, {Stetson}, \& {Schwarzenberg-Czerny}}]{Hartman2006}
{Hartman}, J.~D., {Bersier}, D., {Stanek}, K.~Z., {Beaulieu}, J., {Kaluzny},
  J., {Marquette}, J., {Stetson}, P.~B., \& {Schwarzenberg-Czerny}, A. 2006,
  \mnras, 371, 1405

\bibitem[{{Hiltner}(1960)}]{Hiltner1960}
{Hiltner}, W.~A. 1960, \apj, 131, 163

\bibitem[{{Huxor} {et~al.}(2009){Huxor}, {Ferguson}, {Barker}, {Tanvir},
  {Irwin}, {Chapman}, {Ibata}, \& {Lewis}}]{Huxor2009}
{Huxor}, A., {Ferguson}, A.~M.~N., {Barker}, M.~K., {Tanvir}, N.~R., {Irwin},
  M.~J., {Chapman}, S.~C., {Ibata}, R., \& {Lewis}, G. 2009, \apjl, 698, L77

\bibitem[{{Jester} {et~al.}(2005){Jester}, {Schneider}, {Richards}, {Green},
  {Schmidt}, {Hall}, {Strauss}, {Vanden Berk}, {Stoughton}, {Gunn},
  {Brinkmann}, {Kent}, {Smith}, {Tucker}, \& {Yanny}}]{Jester2005}
{Jester}, S., {Schneider}, D.~P., {Richards}, G.~T., {Green}, R.~F., {Schmidt},
  M., {Hall}, P.~B., {Strauss}, M.~A., {Vanden Berk}, D.~E., {Stoughton}, C.,
  {Gunn}, J.~E., {Brinkmann}, J., {Kent}, S.~M., {Smith}, J.~A., {Tucker},
  D.~L., \& {Yanny}, B. 2005, \aj, 130, 873

\bibitem[{{Kang} {et~al.}(2009){Kang}, {Bianchi}, \& {Rey}}]{Kang2009}
{Kang}, Y., {Bianchi}, L., \& {Rey}, S. 2009, \apj, 703, 614

\bibitem[{{Kotulla} {et~al.}(2009){Kotulla}, {Fritze}, {Weilbacher}, \&
  {Anders}}]{Galev}
{Kotulla}, R., {Fritze}, U., {Weilbacher}, P., \& {Anders}, P. 2009, \mnras,
  396, 462

\bibitem[{{Lamers} {et~al.}(2005){Lamers}, {Gieles}, {Bastian}, {Baumgardt},
  {Kharchenko}, \& {Portegies Zwart}}]{Lamers2005}
{Lamers}, H.~J.~G.~L.~M., {Gieles}, M., {Bastian}, N., {Baumgardt}, H.,
  {Kharchenko}, N.~V., \& {Portegies Zwart}, S. 2005, \aap, 441, 117

\bibitem[{{Larsen}(2002)}]{Larsen2002}
{Larsen}, S.~S. 2002, \aj, 124, 1393

\bibitem[{{Lata} {et~al.}(2002){Lata}, {Pandey}, {Sagar}, \&
  {Mohan}}]{Lata2002}
{Lata}, S., {Pandey}, A.~K., {Sagar}, R., \& {Mohan}, V. 2002, \aap, 388, 158

\bibitem[{{McConnachie} {et~al.}(2009){McConnachie}, {Irwin}, {Ibata},
  {Dubinski}, {Widrow}, {Martin}, {C{\^o}t{\'e}}, {Dotter}, {Navarro},
  {Ferguson}, {Puzia}, {Lewis}, {Babul}, {Barmby}, {Bienaym{\'e}}, {Chapman},
  {Cockcroft}, {Collins}, {Fardal}, {Harris}, {Huxor}, {Mackey},
  {Pe{\~n}arrubia}, {Rich}, {Richer}, {Siebert}, {Tanvir}, {Valls-Gabaud}, \&
  {Venn}}]{McConnachie2009}
{McConnachie}, A.~W., {Irwin}, M.~J., {Ibata}, R.~A., {Dubinski}, J., {Widrow},
  L.~M., {Martin}, N.~F., {C{\^o}t{\'e}}, P., {Dotter}, A.~L., {Navarro},
  J.~F., {Ferguson}, A.~M.~N., {Puzia}, T.~H., {Lewis}, G.~F., {Babul}, A.,
  {Barmby}, P., {Bienaym{\'e}}, O., {Chapman}, S.~C., {Cockcroft}, R.,
  {Collins}, M.~L.~M., {Fardal}, M.~A., {Harris}, W.~E., {Huxor}, A., {Mackey},
  A.~D., {Pe{\~n}arrubia}, J., {Rich}, R.~M., {Richer}, H.~B., {Siebert}, A.,
  {Tanvir}, N., {Valls-Gabaud}, D., \& {Venn}, K.~A. 2009, \nat, 461, 66

\bibitem[{{Mieske} {et~al.}(2010){Mieske}, {Jord{\'a}n}, {C{\^o}t{\'e}},
  {Peng}, {Ferrarese}, {Blakeslee}, {Mei}, {Baumgardt}, {Tonry}, {Infante}, \&
  {West}}]{Mieske2010}
{Mieske}, S., {Jord{\'a}n}, A., {C{\^o}t{\'e}}, P., {Peng}, E.~W., {Ferrarese},
  L., {Blakeslee}, J.~P., {Mei}, S., {Baumgardt}, H., {Tonry}, J.~L.,
  {Infante}, L., \& {West}, M.~J. 2010, \apj, 710, 1672

\bibitem[{{Nantais} {et~al.}(2010){Nantais}, {Huchra}, {Barmby}, \&
  {Olsen}}]{Nantais2010}
{Nantais}, J.~B., {Huchra}, J.~P., {Barmby}, P., \& {Olsen}, K.~A.~G. 2010,
  \aj, 139, 1178

\bibitem[{{Park} \& {Lee}(2007)}]{ParkLee2007}
{Park}, W. \& {Lee}, M.~G. 2007, \aj, 134, 2168

\bibitem[{{Peacock} {et~al.}(2009){Peacock}, {Maccarone}, {Waters}, {Kundu},
  {Zepf}, {Knigge}, \& {Zurek}}]{Peacock2009}
{Peacock}, M.~B., {Maccarone}, T.~J., {Waters}, C.~Z., {Kundu}, A., {Zepf},
  S.~E., {Knigge}, C., \& {Zurek}, D.~R. 2009, \mnras, 392, L55

\bibitem[{{Putman} {et~al.}(2009){Putman}, {Peek}, {Muratov}, {Gnedin}, {Hsu},
  {Douglas}, {Heiles}, {Stanimirovic}, {Korpela}, \& {Gibson}}]{Putman2009}
{Putman}, M.~E., {Peek}, J.~E.~G., {Muratov}, A., {Gnedin}, O.~Y., {Hsu}, W.,
  {Douglas}, K.~A., {Heiles}, C., {Stanimirovic}, S., {Korpela}, E.~J., \&
  {Gibson}, S.~J. 2009, \apj, 703, 1486

\bibitem[{{Regan} \& {Vogel}(1994)}]{ReganandVogel1994}
{Regan}, M.~W. \& {Vogel}, S.~N. 1994, \apj, 434, 536

\bibitem[{{Renzini} \& {Buzzoni}(1986)}]{Renzini1986}
{Renzini}, A. \& {Buzzoni}, A. 1986, in Astrophysics and Space Science Library,
  Vol. 122, Spectral Evolution of Galaxies, ed. {C.~Chiosi \& A.~Renzini},
  195--231

\bibitem[{{Robitaille} \& {Whitney}(2010)}]{Robitaille2010}
{Robitaille}, T.~P. \& {Whitney}, B.~A. 2010, \apjl, 710, L11

\bibitem[{{San Roman} {et~al.}(2009){San Roman}, {Sarajedini}, {Garnett}, \&
  {Holtzman}}]{Sanromanetal2009}
{San Roman}, I., {Sarajedini}, A., {Garnett}, D.~R., \& {Holtzman}, J.~A. 2009,
  \apj, 699, 839

\bibitem[{{Sarajedini} {et~al.}(2007){Sarajedini}, {Barker}, {Geisler},
  {Harding}, \& {Schommer}}]{SarajediniBarker2007}
{Sarajedini}, A., {Barker}, M.~K., {Geisler}, D., {Harding}, P., \& {Schommer},
  R. 2007, \aj, 133, 290

\bibitem[{{Sarajedini} {et~al.}(2000){Sarajedini}, {Geisler}, {Schommer}, \&
  {Harding}}]{Sarajedini2000}
{Sarajedini}, A., {Geisler}, D., {Schommer}, R., \& {Harding}, P. 2000, \aj,
  120, 2437

\bibitem[{{Sarajedini} \& {Mancone}(2007)}]{SarajediniandMancone2007}
{Sarajedini}, A. \& {Mancone}, C.~L. 2007, \aj, 134, 447

\bibitem[{{Stetson}(1987)}]{Stetson1987}
{Stetson}, P.~B. 1987, \pasp, 99, 191

\bibitem[{{Stetson}(1994)}]{Stetson1994}
---. 1994, \pasp, 106, 250

\bibitem[{{Sweigart} {et~al.}(1990){Sweigart}, {Greggio}, \&
  {Renzini}}]{Sweigart1990}
{Sweigart}, A.~V., {Greggio}, L., \& {Renzini}, A. 1990, \apj, 364, 527

\bibitem[{{Tucker} {et~al.}(2006){Tucker}, {Kent}, {Richmond}, {Annis},
  {Smith}, {Allam}, {Rodgers}, {Stute}, {Adelman-McCarthy}, {Brinkmann}, {Doi},
  {Finkbeiner}, {Fukugita}, {Goldston}, {Greenway}, {Gunn}, {Hendry}, {Hogg},
  {Ichikawa}, {Ivezi{\'c}}, {Knapp}, {Lampeitl}, {Lee}, {Lin}, {McKay},
  {Merrelli}, {Munn}, {Neilsen}, {Newberg}, {Richards}, {Schlegel},
  {Stoughton}, {Uomoto}, \& {Yanny}}]{Tucker2006}
{Tucker}, D.~L., {Kent}, S., {Richmond}, M.~W., {Annis}, J., {Smith}, J.~A.,
  {Allam}, S.~S., {Rodgers}, C.~T., {Stute}, J.~L., {Adelman-McCarthy}, J.~K.,
  {Brinkmann}, J., {Doi}, M., {Finkbeiner}, D., {Fukugita}, M., {Goldston}, J.,
  {Greenway}, B., {Gunn}, J.~E., {Hendry}, J.~S., {Hogg}, D.~W., {Ichikawa},
  S., {Ivezi{\'c}}, {\v Z}., {Knapp}, G.~R., {Lampeitl}, H., {Lee}, B.~C.,
  {Lin}, H., {McKay}, T.~A., {Merrelli}, A., {Munn}, J.~A., {Neilsen}, Jr.,
  E.~H., {Newberg}, H.~J., {Richards}, G.~T., {Schlegel}, D.~J., {Stoughton},
  C., {Uomoto}, A., \& {Yanny}, B. 2006, Astronomische Nachrichten, 327, 821

\bibitem[{{Vilardell} {et~al.}(2010){Vilardell}, {Ribas}, {Jordi},
  {Fitzpatrick}, \& {Guinan}}]{Vilardell2010}
{Vilardell}, F., {Ribas}, I., {Jordi}, C., {Fitzpatrick}, E.~L., \& {Guinan},
  E.~F. 2010, \aap, 509, A70+

\bibitem[{{Whitmore} \& {Schweizer}(1995)}]{Whitmore1995}
{Whitmore}, B.~C. \& {Schweizer}, F. 1995, \aj, 109, 960

\bibitem[{{Zloczewski} {et~al.}(2008){Zloczewski}, {Kaluzny}, \&
  {Hartman}}]{Zloczewskietal2008}
{Zloczewski}, K., {Kaluzny}, J., \& {Hartman}, J. 2008, Acta Astronomica, 58,
  23

\end{thebibliography}




\begin{deluxetable}{rcccccccccccc}
\tabletypesize{\scriptsize}
\rotate
\tablewidth{0pt}
\tablecaption{Extended Source Catalog}
\tablehead{
\colhead{Id}	& \colhead{R.A. (J2000.0)}	&
\colhead{Decl. (J2000.0)}	& 
\colhead{g}  & \colhead{(u-g)}   & \colhead{(g-r)}  & \colhead{(g-i)}   & \colhead{(g-z)}  &
\colhead{Ellipticity}   &
\colhead{FWHM ('')}  & \colhead{Stellarity} & 
\colhead{Alter. Id\tablenotemark{a}}   &
\colhead{Classification\tablenotemark{b}}  }
\startdata

     1  & 1 31 33.13  &31 04 05.65  &  \nodata  &  \nodata  &  \nodata  &  \nodata  &  \nodata  &  0.23  &  0.88  &  0.98  & \nodata  & -1 \\
     2  & 1 31 33.39  &30 38 07.97  &  \nodata  &  \nodata  &  \nodata  &  \nodata  &  \nodata  &  0.53  &  1.83  &  0.02  & \nodata  & -1 \\
     3  & 1 31 33.88  &30 58 40.34  &  \nodata  &  \nodata  &  \nodata  &  \nodata  &  \nodata  &  0.37  &  1.58  &  0.01  & \nodata  & -1 \\
     4  & 1 31 33.96  &31 09 09.60  &  \nodata  &  \nodata  &  \nodata  &  \nodata  &  \nodata  &  0.19  &  1.46  &  0.22  & \nodata  & -1 \\
     5  & 1 31 34.05  &31 07 12.55  &  \nodata  &  \nodata  &  \nodata  &  \nodata  &  \nodata  &  0.07  &  1.21  &  0.07  & \nodata  & -1 \\
     6  & 1 31 34.07  &30 40 58.22  &  \nodata  &  \nodata  &  \nodata  &  \nodata  &  \nodata  &  0.06  &  0.83  &  0.93  & \nodata  & -1 \\
     7  & 1 31 34.21  &30 36 09.61  &  \nodata  &  \nodata  &  \nodata  &  \nodata  &  \nodata  &  0.42  &  1.04  &  0.42  & \nodata  & -1 \\
     8  & 1 31 34.36  &31 00 37.44  &  \nodata  &  \nodata  &  \nodata  &  \nodata  &  \nodata  &  0.13  &  1.11  &  0.53  & \nodata  & -1 \\
     9  & 1 31 34.52  &30 53 13.78  &  \nodata  &  \nodata  &  \nodata  &  \nodata  &  \nodata  &  0.39  &  0.00  &  0.35  & \nodata  & -1 \\
    10  & 1 31 34.62  &30 39 10.50  &  \nodata  &  \nodata  &  \nodata  &  \nodata  &  \nodata  &  0.38  &  1.57  &  0.00  & \nodata  & -1 \\
    11  & 1 31 34.67  &30 20 06.59  &  18.464   &   0.950   &   0.750   &   0.932   &  \nodata  &  0.16  &  0.86  &  0.98  & \nodata  &  2 \\
    12  & 1 31 34.78  &31 01 35.22  &  \nodata  &  \nodata  &  \nodata  &  \nodata  &  \nodata  &  0.17  &  1.01  &  0.67  & \nodata  & -1 \\
    13  & 1 31 35.13  &30 17 42.01  &  \nodata  &  \nodata  &  \nodata  &  \nodata  &  \nodata  &  0.77  &  1.07  &  0.05  & \nodata  & -1 \\
    14  & 1 31 35.20  &30 48 11.67  &  \nodata  &  \nodata  &  \nodata  &  \nodata  &  \nodata  &  0.28  &  0.99  &  0.09  & \nodata  & -1 \\
    15  & 1 31 35.28  &30 32 39.49  &  \nodata  &  \nodata  &  \nodata  &  \nodata  &  \nodata  &  0.45  &  1.28  &  0.01  & \nodata  & -1 \\
    16  & 1 31 35.49  &30 45 21.35  &  \nodata  &  \nodata  &  \nodata  &  \nodata  &  \nodata  &  0.21  &  1.33  &  0.00  & \nodata  & -1 \\
    17  & 1 31 35.57  &30 27 44.69  &  \nodata  &  \nodata  &  \nodata  &  \nodata  &  \nodata  &  0.24  &  1.11  &  0.56  & \nodata  & -1 \\
    18  & 1 31 35.59  &30 27 55.07  &  \nodata  &  \nodata  &  \nodata  &  \nodata  &  \nodata  &  0.31  &  1.44  &  0.04  & \nodata  & -1 \\
    19  & 1 31 35.67  &31 00 10.35  &  \nodata  &  \nodata  &  \nodata  &  \nodata  &  \nodata  &  0.24  &  1.28  &  0.11  & \nodata  & -1 \\
    20  & 1 31 35.69  &31 07 25.79  &  \nodata  &  \nodata  &  \nodata  &  \nodata  &  \nodata  &  0.26  &  0.95  &  0.04  & \nodata  & -1 \\
    21  & 1 31 35.70  &30 17 19.51  &   20.107  &    0.957  &   0.453   &   0.625   &   0.704   &  0.19  &  0.89  &  0.93  & \nodata  &  2 \\
    22  & 1 31 35.71  &30 36 09.99  &  \nodata  &  \nodata  &  \nodata  &  \nodata  &  \nodata  &  0.15  &  1.24  &  0.07  & \nodata  & -1 \\
    23  & 1 31 35.89  &30 52 54.88  &  \nodata  &  \nodata  &  \nodata  &  \nodata  &  \nodata  &  0.22  &  1.22  &  0.38  & \nodata  & -1 \\
    24  & 1 31 35.98  &30 23 53.63  &  \nodata  &  \nodata  &  \nodata  &  \nodata  &  \nodata  &  0.28  &  2.57  &  0.06  & \nodata  & -1 \\
    25  & 1 31 36.03  &31 05 26.83  &  \nodata  &  \nodata  &  \nodata  &  \nodata  &  \nodata  &  0.43  &  1.55  &  0.00  & \nodata  & -1 \\
    26  & 1 31 36.05  &30 53 42.02  &  \nodata  &  \nodata  &  \nodata  &  \nodata  &  \nodata  &  0.32  &  1.29  &  0.29  & \nodata  & -1 \\
    27  & 1 31 36.16  &31 06 04.95  &  \nodata  &  \nodata  &  \nodata  &  \nodata  &  \nodata  &  0.21  &  0.89  &  0.89  & \nodata  & -1 \\
 
\enddata
\tablecomments{Table 1 is published in its entirety in the electronic edition of the Journal. The complete table also includes the original CFHT filters (u*g'r'i'z') magnitudes and errors for the candidate star clusters (SC).   \\Units of RA are hours, minutes, and seconds, and
units of Dec are degrees, arcminutes, and arcseconds.}
\tablenotetext{a}{Identification Number in \cite{SarajediniandMancone2007}}
\tablenotetext{b}{Proposed classification: -1 galaxy, 0 unknown extended object, 1 candidate SC, 2 highly probable SC, and 3 confirmed SC (based on SM catalog)}
\label{Table1}
\end{deluxetable}

\begin{deluxetable}{rrcccc}
\tabletypesize{\small}
\tablewidth{0pt}
\tablecaption{Cross Identification with Grossi et al. (2010)}
\tablehead{
\colhead{Id (Us)}	& \colhead{Id (Grossi)}	&
\colhead{log Age\tablenotemark{a}}	& 
\colhead{A$_{v}$\tablenotemark{a}}  & \colhead{E(B-V)\tablenotemark{a}}   & \colhead{Classification\tablenotemark{b}}}
\startdata
  1742  &     VGHC~2-84   & 6.51$\pm$0.14   & 0.58$\pm$0.11   & 0.26$\pm$0.03   & 3  \\
   735  &          C400   & 6.65$\pm$0.15   & 0.52$\pm$0.17   & 0.23$\pm$0.07   & 1  \\
  1144  &         C129a   & 6.35$\pm$0.21   & 0.93$\pm$0.09   & 0.39$\pm$0.02   & 1  \\
  1084  &          C121   & 6.36$\pm$0.13   & 0.93$\pm$0.06   & 0.41$\pm$0.01   & 2  \\
   970  &         B0221   & 6.99$\pm$0.04   & 0.40$\pm$0.02   & 0.16$\pm$0.01   & 2  \\
   760  &     LGC~HII~3   & 6.43$\pm$0.17   & 0.54$\pm$0.08   & 0.22$\pm$0.02   & 2  \\
   847  &         B0261   & 6.53$\pm$0.16   & 0.56$\pm$0.10   & 0.24$\pm$0.04   & 2  \\
   983  &     MCM00Em24   & 6.37$\pm$0.09   & 0.61$\pm$0.03   & 0.28$\pm$0.01   & 2  \\
  1586  &        B0013c   & 7.16$\pm$0.11   & 0.56$\pm$0.11   & 0.27$\pm$0.09   & 3  \\
   952  &          C403   & 6.49$\pm$0.14   & 0.88$\pm$0.11   & 0.38$\pm$0.03   & 2  \\
\enddata
\tablenotetext{a}{Cluster properties from Grossi et al. 2010.}
\tablenotetext{b}{Proposed classification as in Table 1.}
\label{Table2}
\end{deluxetable}

\newpage

%

%

%

%

%

\end{document}